\documentclass[sigconf]{acmart}

\usepackage{tabularx}
    \newcolumntype{L}{>{\raggedright\arraybackslash}X}
\usepackage[inline,nomargin,index]{fixme}
\fxsetup{status=draft,theme=colorsig,mode=multiuser,inlineface=\itshape,envface=\itshape} % show comments
\FXRegisterAuthor{dm}{adm}{Damon}
\FXRegisterAuthor{rg}{arg}{Rachel}
\FXRegisterAuthor{tl}{atl}{Toby}
\FXRegisterAuthor{cb}{acb}{Cameron}
\FXRegisterAuthor{ig}{aig}{Ian}
\FXRegisterAuthor{kt}{akt}{Kejsi}
\FXRegisterAuthor{vz}{avz}{Victoria}
\FXRegisterAuthor{gs}{ags}{Genesis}
\FXRegisterAuthor{pm}{apm}{Pulak}
\usepackage{siunitx}

\copyrightyear{2022}
\acmYear{2022}
\setcopyright{acmlicensed}\acmConference[WWW '22]{Proceedings of the ACM
Web Conference 2022}{April 25--29, 2022}{Virtual Event, Lyon, France}
\acmBooktitle{Proceedings of the ACM Web Conference 2022 (WWW '22), April
25--29, 2022, Virtual Event, Lyon, France}
\acmPrice{15.00}
\acmDOI{10.1145/3485447.3512142}
\acmISBN{978-1-4503-9096-5/22/04}

\setcopyright{none}
\renewcommand\footnotetextcopyrightpermission[1]{} % removes footnote with conference information in first column
\begin{document}

\title{Conspiracy Brokers: Understanding the Monetization of YouTube Conspiracy Theories}

\author{Cameron Ballard} 
\affiliation{
    \institution{New York University}
    \streetaddress{370 Jay Street}
    \city{Brooklyn}
    \state{New York}
    \country{USA}
    \postcode{11201}
}
\author{Ian Goldstein}
\affiliation{
    \institution{New York University}
    \streetaddress{370 Jay Street}
    \city{Brooklyn}
    \state{New York}
    \country{USA}
    \postcode{11201}
}
\author{Pulak Mehta}
\affiliation{
    \institution{New York University}
    \streetaddress{370 Jay Street}
    \city{Brooklyn}
    \state{New York}
    \country{USA}
    \postcode{11201}
}
\author{Genesis Smothers} 
\affiliation{
    \institution{New York University}
    \streetaddress{370 Jay Street}
    \city{Brooklyn}
    \state{New York}
    \country{USA}
    \postcode{11201}
}
\author{Kejsi Take} 
\affiliation{
    \institution{New York University}
    \streetaddress{370 Jay Street}
    \city{Brooklyn}
    \state{New York}
    \country{USA}
    \postcode{11201}
}
\author{Victoria Zhong}
\orcid{0000-0001-9300-3005}
\affiliation{
    \institution{New York University}
    \streetaddress{370 Jay Street}
    \city{Brooklyn}
    \state{New York}
    \country{USA}
    \postcode{11201}
}
\author{Rachel Greenstadt} 
\affiliation{
    \institution{New York University}
    \streetaddress{370 Jay Street}
    \city{Brooklyn}
    \state{New York}
    \country{USA}
    \postcode{11201}
}
\author{Tobias Lauinger}
\orcid{0000-0002-5779-0643}
\affiliation{
    \institution{New York University}
    \streetaddress{370 Jay Street}
    \city{Brooklyn}
    \state{New York}
    \country{USA}
    \postcode{11201}
}
\author{Damon McCoy} 
\affiliation{
    \institution{New York University}
    \streetaddress{370 Jay Street}
    \city{Brooklyn}
    \state{New York}
    \country{USA}
    \postcode{11201}
}

% \author{Ben Trovato}
% \authornote{Both authors contributed equally to this research.}
% \email{trovato@corporation.com}
% \orcid{1234-5678-9012}
% \author{G.K.M. Tobin}
% \authornotemark[1]
% \email{webmaster@marysville-ohio.com}
% \affiliation{%
%   \institution{Institute for Clarity in Documentation}
%   \streetaddress{P.O. Box 1212}
%   \city{Dublin}
%   \state{Ohio}
%   \country{USA}
%   \postcode{43017-6221}
% }

\begin{abstract}
    Conspiracy theories are increasingly a subject of research interest as society grapples with their rapid growth in areas such as politics or public health. Previous work has established YouTube as one of the most popular sites for people to host and discuss different theories. In this paper, we present an analysis of monetization methods of conspiracy theorist YouTube creators and the types of advertisers potentially targeting this content. We collect 184,218 ad impressions from 6,347 unique advertisers found on conspiracy-focused channels and mainstream YouTube content. We classify the ads into business categories and compare their prevalence between conspiracy and mainstream content. We also identify common offsite monetization methods. In comparison with mainstream content, conspiracy videos had similar levels of ads from well-known brands, but an almost eleven times higher prevalence of likely predatory or deceptive ads. Additionally, we found that conspiracy channels were more than twice as likely as mainstream channels to use offsite monetization methods, and 53\% of the demonetized channels we observed were linking to third-party sites for alternative monetization opportunities. Our results indicate that conspiracy theorists on YouTube had many potential avenues to generate revenue, and that predatory ads were more frequently served for conspiracy videos.

\end{abstract}

%%
%% The code below is generated by the tool at http://dl.acm.org/ccs.cfm.
%% Please copy and paste the code instead of the example below.
%%


%%
%% Keywords. The author(s) should pick words that accurately describe
%% the work being presented. Separate the keywords with commas.

%\settopmatter{printacmref=false}
%\setcopyright{none}
%\renewcommand\footnotetextcopyrightpermission[1]{}

%\pagestyle{plain}

% \acmConference[The Web Conference '22]{}

%%
%% This command processes the author and affiliation and title
%% information and builds the first part of the formatted document.
\settopmatter{printfolios=true}
\maketitle
\renewcommand{\shortauthors}{Ballard et al.}

\section{Introduction}

\label{sec:introduction}
As political and anti-vaccination conspiracy theories proliferate across the web, conspiracy theories have increasingly been a subject of academic and journalistic research. Perhaps the most well known, Flat Earth, has gathered a global following across various social media platforms by blending elements of science denial, religious fundamentalism, and antisemitism into a powerful vehicle for the spread of misinformation. Participants in a Texas Tech University survey of individuals at a Flat Earth convention near unanimously pointed to YouTube as their entry into the community~\cite{olshansky18}. The vast majority also reported being introduced to more conspiracies through YouTube after finding the Flat Earth community~\cite{olshansky18}.
Outside of Flat Earth Theory, the prevalence of COVID-19 anti-vaccination conspiracies, and the role QAnon played in the US Capitol Riot on January 6th, 2021~\cite{qanonriots,qanonriots2} have demonstrated the real danger that these theories can pose to individuals and society.

Alongside the societal impacts of theories like QAnon, viewers of conspiracy theories could become the target of predatory advertising practices. While the exact traits that contribute to an individual believing in conspiracy ideology remain unclear~\cite{notgullible}, studies have established a link between belief in conspiracy theories and gullibility or lack of critical thinking~\cite{gullible,criticalthinking}. Just as online scammers might use poorly constructed emails to identify gullible victims and avoid ``false positives''~\cite{mai18,nigerianscammers}, the targeting of viewers of conspiracy theories might allow predatory advertisers to increase the success of their campaigns, intentional or not.

In this paper, we investigate how ads shown on conspiracy content on YouTube differ from ads on mainstream YouTube videos, and whether they advertise a higher rate of dubious products or services. 
As a second research question, we investigate how creators of conspiracy content monetize their videos, e.g., through YouTube ads and offsite mechanisms such as donation links. Understanding the available monetization strategies for publishers of conspiracy theories is key to addressing the infrastructure and incentives that allow and even encourage their spread across the internet.

Prior research has identified online advertising as a lucrative source of revenue for fake news publishers~\cite{braun19,alba17,funke19,espinoza20}, but did not specifically analyze conspiracy theories or YouTube. 
To the best of our knowledge, our study is the first attempt to understand the monetization strategies of conspiracy theorists on YouTube and the advertisers that profit from and indirectly finance them.

To answer our two research questions, we regularly visited a set of conspiracy videos and compared the frequency and types of ads with a control set of mainstream videos over a period of 4.5 months.
Our results indicate that the rate of ads on conspiracy videos was 16\% of that on mainstream videos.
In part, this seems to be due to videos being ``demonetized'' when their content is not considered advertiser-friendly.
Conversely, conspiracy videos were 2.3 times more likely to use at least one offsite monetization method such as donation links, and 5.2 times more likely to use multiple.
When an ad was shown, it was 10.7 times as likely as control to be predatory.
Overall, we identify an ecosystem that gives content creators opportunities to monetize conspiracy videos on and off YouTube, and assists predatory advertisers to target their viewers. 

Our work makes the following contributions:
\begin{itemize}
    \item We are the first to study the monetization strategies of conspiracy theorists on YouTube. 
    \item We show that conspiracy videos have fewer, lower-quality, and more predatory ads than mainstream videos on YouTube.
    \item We find that conspiracy channels are much more likely to use offsite monetization strategies than mainstream creators. 
\end{itemize}

\section{Background \& Related Work}
\label{sec:background}
Specifically defining what is and is not a conspiracy theory can be difficult. Oxford Dictionary defines it as ``a belief that some covert but influential organization is responsible for a circumstance or event.'' However, individual theories can vary in intensity and specificity---such as a belief in the existence of an indistinct global elite versus the belief in a council of ten people who control the entire world---and often incorporate events or theories from many sources, including politics, pop culture, and religion.

Conspiracies on YouTube are dynamic and greatly varied, adapt quickly to current events, and include topics such as the QAnon conspiracy, New World Order, Galactic Federation, vaccine skepticism, COVID denial, Jeffrey Epstein's suicide, biblical predictions, higher consciousness, aliens, UFOs, and the deep state. Topic-modelling on conspiracy videos found alternative science and history, prophecies and cults, and political conspiracies as the most common topics~\cite{faddoul20}. 

In practice, these conspiracies develop, fracture, and merge with other theories, becoming loose headings for the grouping of various misinformation. As an example, Garry et al. traced the origins QAnon to the pizzagate conspiracy---or the belief that democrats were running a child sex-trafficking ring out of a pizza parlor~\cite{Qanonderad}. Pizzagate began in response to the release of Clinton's emails in 2016, and was definitively debunked when one of its adherents arrived at the restaurant with guns, demanded access, and found no evidence of sex-trafficking. Rather than giving up the theory, the more dedicated believers argued that something much more sinister was happening in the U.S.\ government. This suspicion led to the creation of increasingly outlandish conspiracies, such as the belief that politicians and celebrities were drinking the blood of tortured children to gain immortality. QAnon adopted a number of these theories, ultimately priming a ready group of believers for ``The Big Lie'' and anti-vaccination COVID misinformation\cite{qanonrise}, and significantly contributing to the storming of the U.S.\ capitol building on January 6th, 2021~\cite{qanonriots,qanonriots2,qanonsiege}. While the growth of QAnon was not limited to YouTube, the site is one of the most popular mainstream digital spaces for conspiracy viewers to congregate, discuss, and find new theories\cite{olshansky18}, making it well-suited to creators looking to monetize their content. Understanding the financial ecosystem around these conspiracy videos contributes to better understanding the monetization of misinformation as a whole.

\subsection{Advertising on YouTube}
We identified three distinct types of advertising on YouTube.
Most noticeable are the video advertisements served before or during a video.
Along with these advertisements, YouTube places image advertisements
on the video sidebar.
The sidebar advertisements we observed were almost always from the same
source as the video advertisement.
YouTube also runs simpler text and image ``banner'' ads in the
middle of videos that do not interrupt viewing.

\subsubsection{YouTube's Ad Delivery System}
Advertising on YouTube is managed through Google’s digital ads platform.
As part of their ad campaigns, advertisers can upload a combination of text, images, and video.
They can specify on which sites or apps the advertisement can be seen and
how the ad should be targeted.
Advertisers also specify the amount of money they are willing to spend
per interaction with an ad, referred to as a bid.
Through a combination of content restriction, bidding, and
personalization, Google ultimately determines where and when these ads
will be served. 

Google allows advertisers to target ads based on audience characteristics (e.g., personal demographics or interests of the viewer) and ad context (i.e., where the ad is shown).
Because of the variety of targeting methods based on audience
characteristics, and the difficulty of emulating realistic profiles of
users interested in conspiracy videos, we focus our study on context-based
targeting.
In that category, Google allows targeting based on specific video, channel,
or website placements, topic and keywords associated with content, or
specific devices (e.g. only show on iPhones).
Additionally, ad delivery systems typically ``optimize'' further within
the targeting criteria specified in a campaign by selecting users or
advertising opportunities that the platform considers most likely to
result in an interaction with the ad.
By monitoring a specific subset of content with ad personalization
disabled, we can observe the combined effect of content-based targeting and (non-personalized) delivery optimization, but unfortunately cannot distinguish between them.

\subsubsection{(De)Monetization and the Partner Program}
Several factors determine if YouTube videos can be monetized
with ads delivered by the platform.
The ``YouTube Partner Program'' allows creators to derive revenue from
ads shown on their content when their channel follows all community
guidelines, has over 1000~subscribers, and has been viewed for over
4000~hours in the past 12~months~\cite{Ytpartner}.
Alternatively, creators can monetize their videos using AdSense,
Google's ad delivery system for the broader internet~\cite{Nelsonadsense}.
Even when a creator does not participate in any of these programs, YouTube
may choose to run ads on the video, retaining all of the
revenue~\cite{Policychange}.

Videos that violate YouTube's policies or contain content that is not
``advertiser friendly'' (e.g., nudity, violence, hateful
content, or sensitive events~\cite{Ytguidelines,Ytadfriendliness}) can be flagged for review and ``demonetized,''
preventing ads from being served to that video.
YouTube conducts regular automatic and manual reviews of monetized channels~\cite{Ytguidelines}. 
Creators may also preemptively demonetize their videos.

\subsection{Off-Platform Monetization Opportunities}
YouTube creators often leverage additional off-platform methods to
improve revenue generation from their content.
Websites such as Patreon allow users to support creators directly.
Other websites like GoFundMe and PayPal allow creators to accept direct
donations, either for a specific cause or their channel as a whole.
Creators may also enter into sponsorship agreements with certain
companies where they actively promote a product in their videos in
return for monetary support.
Additionally, YouTube allows creators to advertise merchandise below
their videos and link to outside stores.
These monetization opportunities allow creators to generate income from
their content even after being disallowed from
running ads.

\subsection{Related Work}
Related to our work are studies of conspiracy and misinformation on
social media, and measurements of online advertising.

\noindent\textbf{Online Misinformation and Conspiracies.} 
A lot of work has gone into understanding the spread of
online extremism and misinformation across social networks,
mainly on large text-based networks such as Twitter and
Facebook~\cite{Kumar18,Vicario16,Allcott19}.  
%An earlier study in 2016 found that Facebook facilitated the increased
%spread of misinformation around conspiracy theories and scientific topics
%\cite{Vicario16}.
%This activity was confirmed later by a study which found activity around
%false information on Facebook and Twitter rose steadily from 2015--2016,
%before falling some on Facebook \cite{Allcott19}.
Researchers~\cite{Bradshaw18,Asmolov18} and investigators~\cite{mueller19}
have also studied foreign influence and governmental social media
campaigns.
However, YouTube is beginning to receive more attention, with several 
recent studies examining misinformation and hate speech shared on 
the site \cite{hussein20,papadamou21,tomlein21,jagtap21}.
%Given the commonly political nature of misinformation in 2016, a lot of
%work has focused on foreign influence and governmental social media
%campaigns \cite{Bradshaw18} \cite{Asmolov18}.
%Online misinformation even received official governmental attention, with
%Russian social media campaigns extensively discussed in Robert Mueller's
%report to congress on the 2016 election~\cite{mueller19}.
%
%Along with active foreign influence campaigns came a proliferation of
%user-driven misinformation, including content from alternative news
%networks and other creators that is discovered, reproduced, and shared by
%other social media users.

Especially after the start of the COVID-19 pandemic and the U.S.\ Capitol
Riots of January 6th, 2021, science denial, conspiracy theories, and
their crossover with alternative political ideologies have also become the
subject of much misinformation research.
One early example is the Flat Earth Theory, which blends elements of science
denial, religious fundamentalism, antisemitism, and conspiracy theory into
a popular YouTube community~\cite{Paolillo18}.
In a study of a Flat Earth convention, nearly all respondents reported
that they discovered and were convinced by the theory from YouTube videos,
with many pointing to the website as their entry into other online
conspiracies as well~\cite{olshansky18}.
Other work has examined the spread of QAnon content
across social media platforms (including
YouTube)~\cite{Normiefication, PapasavvaVoat, aliapoulios21}. One paper specifically focused on automatically classifying conspiracy videos on YouTube \cite{faddoul20}.
% including Facebook, Twitter, Reddit, Voat, and YouTube
A 2014 study found that exposure to conspiracy content reduced an
individual's willingness to receive
vaccinations~\cite{reuters20,roozenbeek20,kouzy20,jolley14}. 
Other research additionally supports the overlap of conspiracy and
political communities such as the alt-right and radical
left~\cite{ribeiropathways,ledwichzaitsev20}. 
%Perhaps the most obviously damaging online conspiracy, QAnon, saw many
%supporters arrested after the Capitol Riots on January 6th which left
%five people dead \cite{qanonsiege}.
%The COVID-19 pandemic has also seen the proliferation of various
%anti-vaccination misinformation and conspiracy theories.
%Extensive work has already been carried out analyzing the spread and
%impact of some of this content across many networks, with one study in
%2014 finding that exposure to conspiracy content reduced an individual's
%willingness to receive vaccinations \cite{reuters20} \cite{roozenbeek20} \cite{jolley14} \cite{kouzy20}.
%While our study does very little to analyze the actual content of
%conspiracy videos, it is important to note the prevalence of this content
%on YouTube, and its significant impact on real world events. 

%\vlnote{seemingly very related work? https://dl.acm.org/doi/10.1145/3392854} \cbnote{omg}

%\cbnote{also need to cite "A longitudinal analysis of YouTube’s promotion of conspiracy videos"}

\noindent\textbf{Online Advertising.}
Prior work on online advertising systems includes general measurements~\cite{barford14}, or work on
discrimination~\cite{adfisher,ali19,datta18} and
transparency~\cite{edelson20}.
%Online advertising systems have also received more attention in recent
%years, especially the automatic targeting systems.
%Most of this research has been focused on Facebook and Google's ad
%delivery systems since they account for the majority of ads seen online.
%Some studies use the ad delivery system as a starting point for research.
%The Adfisher project explored the accuracy of Google's ad transparency
%measures, and found evidence that targeting can lead to discriminatory
%outcomes \cite{adfisher}.
%The discriminatory effects of ad delivery systems have been further
%supported by later studies of Facebook's
%platform \cite{ali19} \cite{datta18}.
%Others have sought to analyze ad transparency from the user side,
%scraping sites or getting user-generated data for analysis of ad
%libraries.
%Barford crawled websites with different user profiles to observe what ads
%were shown \cite{barford14}.
%Edelson utilized a browser extension \tlerror{this seems incorrect, unsure where it's coming from}
%to collect advertisements seen by
%actual users across Facebook and measured the effectiveness of their
%political ad transparency initiatives \cite{edelson20}.
%Advertising on YouTube has received some attention, but comparatively
%less than Facebook or Google as a whole.
On YouTube, researchers have studied the reach and effectiveness of
political advertising~\cite{ridout10,alujevic14}, the effects of ads on
children~\cite{tan18,vanwesenbeeck20}, and from a business perspective,
the effectiveness of YouTube advertising in attracting new
customers~\cite{rodriguez17,dehghani16}.
%An earlier study examined the reach and effectiveness of political
%advertising on YouTube specifically \cite{ridout10}.
%Vesnic-Alujevic performed a similar analysis of YouTube ads in the
%context of the European Parliament elections in 2010 \cite{alujevic14}.
%Two other studies examined the ads served to children and how the
%children process YouTube ads differently than traditional TV
%advertising \cite{tan18} \cite{vanwesenbeeck20}.
%From a business perspective, Dehghani and Rodriguez both examined the
%effectiveness of YouTube advertising in attracting new
%customers \cite{rodriguez17} \cite{dehghani16}.
Other work has also explored influencer marketing and sponsorship~\cite{rasmussen18},
and disclosure of affiliate marketing~\cite{affiliateMarketingDisclosureMeasurement,affiliateMarketingBrowserExtension}.

Research has also identified potential misuses of online advertising.
Sood laid out strategies used in ``malvertising,'' or the use of online
ads to spread malware~\cite{sood11}, a concept further explored by later
research~\cite{xing15}.
Other studies have identified possible injection attacks through browser
advertising~\cite{thomas15}, and advertisements for software that comes
packaged with additional unwanted software (potentially
unwanted programs, PUP) distributed through pay-per-install (PPI)
advertising~\cite{kotzias16,thomasppi}.
While our study did not identify any advertisements actively distributing malware, we
did observe advertisements for PUP and ads from domains that had been flagged for serving malware in the past. A 2021 study from Zeng et al.\ outlined types of ads that users considered `bad'~\cite{badads}, a significant number of which we encountered in our study.

\noindent\textbf{Monetization of Mis- and Disinformation.}
Recent work has begun to examine the network of funding around
``fake news'' on social media platforms.
With the exception of one academic study on the advertising ecosystem
around misinformation that outlined a ``lucrative incentive structure''
for fake news publishers~\cite{braun19}, so far this line of work has
been mostly journalistic, and shown that advertising remains a significant
source of income for publishers of fake
news~\cite{alba17,funke19,espinoza20}.
To the best of our knowledge, there has been no peer-reviewed work so far
on the types of advertisers shown in the context of conspiracy theory videos,
or the monetization of conspiracy theories or misinformation on YouTube specifically.

\section{Methodology}
\label{sec:methodology}

We sought to understand the methods conspiracy channels use to monetize both on and off of YouTube, and the type of advertisements being targeted at this content. We began by crawling a set of conspiracy videos and a control group of mainstream content, collecting various metadata on each video and channel as well as any advertising information. After collecting advertising data over several months from multiple locations, the most commonly occurring advertisements were classified into different categories of business and product offerings. Finally, we compared the prevalence and types of advertising seen as well as other monetization opportunities leveraged by the creators. 

\subsection{Data Collection}

In order to understand what advertisements are being disproportionately targeted at conspiracy content, we needed information from both conspiracy content and more mainstream content. Using classifications from prior studies, we compiled a list of conspiratorial videos from a smaller group of channels we could confidently say were conspiracy focused, and a larger group of automatically identified channels with a shared audience interested in conspiracy theories. We regularly visited these videos alongside popular YouTube videos with a headless browser.

To make sure the ads we saw were selectively targeted towards conspiratorial content and not the result of site-wide campaigns, we needed a group of generic ads to compare against. To gather these ads, we collected data from a control set of videos pulled from recommended YouTube content. The two data sets and channel sources are discussed in detail in Sections~\ref{sec:controldata} and~\ref{sec:conspiracydata}.

\subsubsection{Crawler Process}
Both of these sets were scraped on virtual machines (VMs) in Iowa and Oregon. Nearly all of the conspiracy channels were in English, thus we did not attempt to collect data outside the U.S.
The VMs ran Windows Server 2019 and crawling was done with Selenium and Google Chrome.
We wanted to minimize any data contamination from tracking that may be performed by Google for the purpose of serving ads, so each data set had an associated VM and YouTube account for collecting advertisements. Crawling was carried out from June 1st to October 13th, 2021.

YouTube normally personalizes the ads that users see. It is unclear how exactly YouTube does personalization, but it likely relies on some collection of user activity from across the web.
Accurately replicating the activity of a real human would be extremely difficult without knowledge of how exactly YouTube serves advertisements. Since we were interested in finding the ads displayed on a particular type of content rather than a subgroup of individuals, we turned off ad personalization for the YouTube accounts used in scraping. According to the targeting criteria information YouTube provides, we were served ads only based on the video we watched, our approximate location, and the time of day. Since location and time of day were still being used for targeting, we collected ads throughout the day and from two different locations, namely Iowa and Oregon.

For each video, we collected associated metadata, the video name and ID, the upload date, the video description and any URLs listed, any content warnings, and removal information if the video was taken down.
If a pre-roll advertisement was seen, the crawler gathered any advertising information and moved on to the next video in the list. If no pre-roll advertisement was seen, the crawler skipped to later in the video and waited a short interval for a banner ad to appear. If no banner ad was found, the crawler continued through the video list. Any sidebar image advertisements were ignored, as they were almost always from the same source as the video advertisement that appeared. This means any visit to a video only collected information for a single advertisement, even though sometimes multiple preroll ads are shown in succession. 
For advertisements, we collected the destination URL of the ad, its type (video or banner), and the targeting information provided by YouTube.

% For the conspiracy data set, the crawler operated off a randomly selected subset of videos to visit. After logging into YouTube, the crawler visited each video in the subset.
% For the control set, the crawler took random walks through recommendations from the homepage. 
Due to the dynamic nature of the website and some built-in crawler protection, the crawler frequently crashed before completing a given list of videos. As such, the crawler was set to relaunch with a new set of 250 videos each hour. By restarting random data collection throughout the day, we were able to collect data from different time periods and without a specific ordering of the videos. On average, each run we visited 190 videos in the conspiracy set and 135 in the control set, with the difference due to increased interaction with the site in the control set often breaking the crawler.

It is important to note that our crawling was not exhaustive. The advertisements we saw do not represent the totality of ads served on YouTube or any specific video. The presence and number of ads on any given video is determined stochastically, but by randomizing video visits, we were able to gain a general view of the types of advertisements running on videos we tracked. We limited the majority of our analysis only to ads that we saw 10 or more times to protect against any noise introduced by the stochastic ad delivery. The implications of this are explored further in Section~\ref{sec:limitations}.

\subsubsection{Control Set}
\label{sec:controldata}

To understand what ads are targeted towards conspiratorial content, we must first establish a control set of ads seen on ``mainstream'' YouTube content. We chose to use YouTube's own recommendation engine to establish what Google views as mainstream content rather than rely on outside analysis. The scraper picked a random video from the homepage of YouTube and then took random walks down the recommendation engine. After visiting anywhere from ten to fifty recommended videos, the crawler returned to the homepage, picked a random video, and proceeded through the recommendations again. Each run of the crawler generated a different set of control videos, and we did not review exactly what content was visited.
Table~\ref{tab:datasets} summarizes the data set.

Because we turned off video personalization and relied on homepage videos, the crawler saw content that YouTube recommends to new users and thinks a majority of people will want to see.
Each walk down the recommendation engine would imply a grouping of similar content, but by frequently restarting the recommendation walks, % and keeping video personalization off,
the crawler was never pigeonholed into one type of content. This results in a broad range of mainstream video and ad content. %, while ensuring we stay within YouTube's most generic and mainstream recommendations. %, establishing some level of ground truth for advertisements seen on mainstream YouTube videos from a variety of topics.
This selection of a control set does not provide a comprehensive view of YouTube's advertising, but it does give us a set of ads that a very large portion of YouTube users would view.
%It also lets us determine what ads were served to conspiracy videos specifically and not site wide. 

\begin{table}[t]
\footnotesize
\begin{tabular}{lrrr}
\toprule
     \textbf{Dataset} & \textbf{Channels} & \textbf{Videos} & \textbf{Ad Impressions} \\
\midrule
  \textbf{Conspiracy} &           818 &      93,449 &         43,379 \\
     \textbf{Control} &         11,912 &       47,847 &        140,839 \\
\bottomrule
\end{tabular}
\caption{Overview of the two data sets.}
\label{tab:datasets}
\end{table}

\subsubsection{Conspiracy Data Sets}
\label{sec:conspiracydata}

Data collection for the conspiracy set was carried out on two different sources of conspiracy videos. For the first, we sought a group of videos %we could confidently say were
focused on %or consistently discussed
conspiratorial or pseudoscientific content. We extracted them from Ledwich and Zaitsev~\cite{ledwichzaitsev20}, who manually labeled many YouTube channels. The paper focused on political channels, but discovered enough conspiratorial YouTube creators to give them a specific category. 
From these 116 channels, we gathered a list of 22,337 videos, stretching back to videos uploaded in 2007.
Using data from RadiTube~\cite{raditube}, we found that 40\% of videos had been removed;
we excluded them from analysis because they do not trigger any ads.

For the second source, we cast a much broader net to encompass the wide range of different conspiracy channels. These came from Clark and Zaitsev~\cite{chan2vec}, who classified channels based on shared subscribers. While this method does not explicitly classify the content itself, it does establish a community of viewers and creators that are interested in similar content, and reports a model agreement score within 1\% of labeler agreement. To generate the list of videos, we selected all channels labelled as ``conspiracy'' or ``qanon'' content and pulled all videos uploaded from January 2020 until May 2021. This method gave us a larger set of 72,343 recent videos from 741 channels, including 38 from the smaller list of videos. 

%\tlnote{Could potentially add here some more characterisation/examples of what type of content is in these conspiracy channels (to reuse the text cut from the Background section/2.1)}

For conspiracy data collection, the total list of videos observed by the crawler remained static after initial collection, and does not include any new videos uploaded after May 28th, 2021. Each run of the crawler randomly selected and ordered a subset of this fixed video list for collection, ensuring we visited individual videos many times and in different orders. To consistently visit all channels in the dataset  we first selected a random channel and then a video from that channel until 250 videos had been chosen.

After collection, we observed that the ads seen from both sources of conspiracy videos and at both geographical crawler locations were very similar. As such, for analysis we combined all conspiracy data from both locations into a single set of conspiracy videos, and compared this to a similarly combined control data set. By combining locations and analysing the most frequently occurring ads, we can minimize the impact of any location-based targeting.

\subsection{Labeling}

We classified the most frequently seen advertisements according to two categorizations. For the first category, \textit{Business Type}, we split the ads into different types of businesses such as traditional merchants or self-improvement ads. For the second, \textit{Content Type}, we split the ads into different content areas, such as food and drink or gaming. Advertisements were labeled based on the page they linked to, not the actual ad shown on YouTube. 

To select the advertisements for labeling, we split them into three subsets: those found only in the control set, only in the conspiracy set, and ads found in both sets. Examining these subsets separately is key to understanding what ads are shown specifically on conspiracy content and not simply site-wide. The top 300 most frequently encountered ads from each of these subsets were selected, as well as 200 ads randomly sampled from those we encountered only 3-5  times. Labeling of advertisement types was carried out iteratively. We performed two rounds of labeling and consolidation of content categories to find distinct and understandable categories. An explanation of the categories chosen can be found in Table~\ref{tab:catsubcat} in the appendix. Ultimately, five labelers were used for our final round of annotation. A smaller number of ads (60) was labeled by all five people, with strong agreement---Fleiss kappa 0.9 for business type and 0.79 for content type. The rest of the labels were divided up randomly and labeled individually. While labeling, the labelers were not informed whether an ad was shown on a conspiracy video so that this knowledge could not affect their chosen label.

\section{Analysis}
\label{sec:analysis}

Overall, we observed 184,218 ad impressions from 
6,347 unique advertisers across all data sets. Advertisers were grouped by the domain of the URL linked to in the ad. %(e.g., \textit{chanel.com/watches} and \textit{chanel.com/jewelry} would be considered the same advertiser, \textit{chanel.com}).
Conspiracy videos (from both geographical locations) saw ads from 3,475 of these advertisers, control from 4,277.
The large majority of advertisements we only encountered a few times; 41\% of advertisers we only observed once, and 68\% less than 5 times (shown in Figures~\ref{fig:adshistogram} and~\ref{fig:adscdf} in the appendix).
To ensure statistical significance, we limit any categorical analysis to the 300 ads most frequently seen only in control, only in conspiracy, and in both data sets, as well as a random sample of 200 other ads. While this subset of 1,100 ads only accounts for 17\% of the observed advertisers, it includes 88\% of all impressions.

Ads differed much more between the control and conspiracy sets than between geographical locations of the crawler. We measured the degree of overlap in the ads using a [0,1]-cosine similarity score on the vector of unique advertisers weighted by observation frequency. On average, content from the same data set (i.e., conspiracy or control) but different locations had a similarity score above 0.594, while content from different data sets but the same location had an average similarity score below 0.276. In other words, the advertisements seen on the conspiracy data sets from the two locations were roughly twice as similar to each other than to the advertisements seen on the control videos. The greater dissimilarity in ads between control and conspiracy implies that while location targeting did occur, YouTube's ad serving system did result in selective delivery of advertisements to conspiracy videos.

\subsection{Demonetization}
In general, advertising was much more common on control videos, appearing in roughly 50\% of visits to any given video. For conspiracy videos where we did see ads, it took an average of 4 visits before the first ad impression. 95\% of videos saw an impression after 10 or fewer visits. 
Despite an average of 11 visits per conspiracy video, we did not observe any ads on 82\% of conspiracy videos. 
These videos might have been demonetized, that is, excluded from advertising because their content is not deemed advertiser friendly. 
In comparison, only 29\% of control videos that we visited more than 10 times did not show advertisements. 
Due to the random delivery of ads and lack of transparency on YouTube's content moderation, precisely measuring demonetization through scraping is difficult. We cannot definitively say a video or channel is demonetized, only that we did not observe ads. However, after a certain number of visits to a channel or video, it is highly probable the cause is demonetization.

Demonetization behavior we found is consistent with YouTube's stated policies. In the conspiracy data set, we observed 348 channels that did not display ads and 467 that did, with 1,059 average visits per channel. Of channels where we saw ads, 182 had videos that did not display ads after thirty or more visits, and 111 after 50+ visits, making it highly probable that YouTube performs channel and video-level demonetization. We also observed a higher rate of video removal for likely demonetized channels, at 17\% for monetized channels and 32\% for demonetized,
suggesting creators who repeatedly violate community guidelines and have their videos removed are more likely to be prevented from running ads. Out of 37 channels that were fully terminated during our collection period, 10 were able to run ads before they were removed from the site. 

The average number of views was also much higher for monetized videos. In the conspiracy set, monetized videos had almost four times as many views as demonetized videos on average (42,776 and 11,811). Part of this may be due to the YouTube Partner Program, which allows channels with a certain amount of monthly watch time to automatically monetize their videos. However, even in the control set, where average views was in the millions---more than enough to cross the Partner Program threshold---the trend held true (6.2 vs.\ 2.9 million views). For control videos we visited 10 or more times, and thus can confidently say were or were not monetized, the difference was even larger at 12.9 and 3.7 million average views. Given the strong financial incentive to recommend videos with ads, the relatively higher amount of views on monetized content suggests YouTube may more often promote content they can serve ads on. 

Demonetization aside, even among the conspiracy videos for which we observed at least one ad, each visit had a roughly 20\% chance to see an ad, compared to almost 50\% in the control set. Banner advertisements caused most of the increased prevalence of ads in control. On conspiracy videos with ads, we saw a preroll advertisement roughly one in every four visits on average, and a banner ad one in every 10. In comparison, control videos with ads saw a preroll advertisement one in every ten, but a banner advertisement more than every other visit. While the mechanism behind this difference is unclear, the result is fewer advertisements served on conspiracy channels, but a higher prevalence of predatory advertisements like those discussed in Section~\ref{sec:badads}.

\subsection{Offsite Monetization Strategies}
\begin{figure}
    \centering
    \includegraphics[width=\columnwidth]{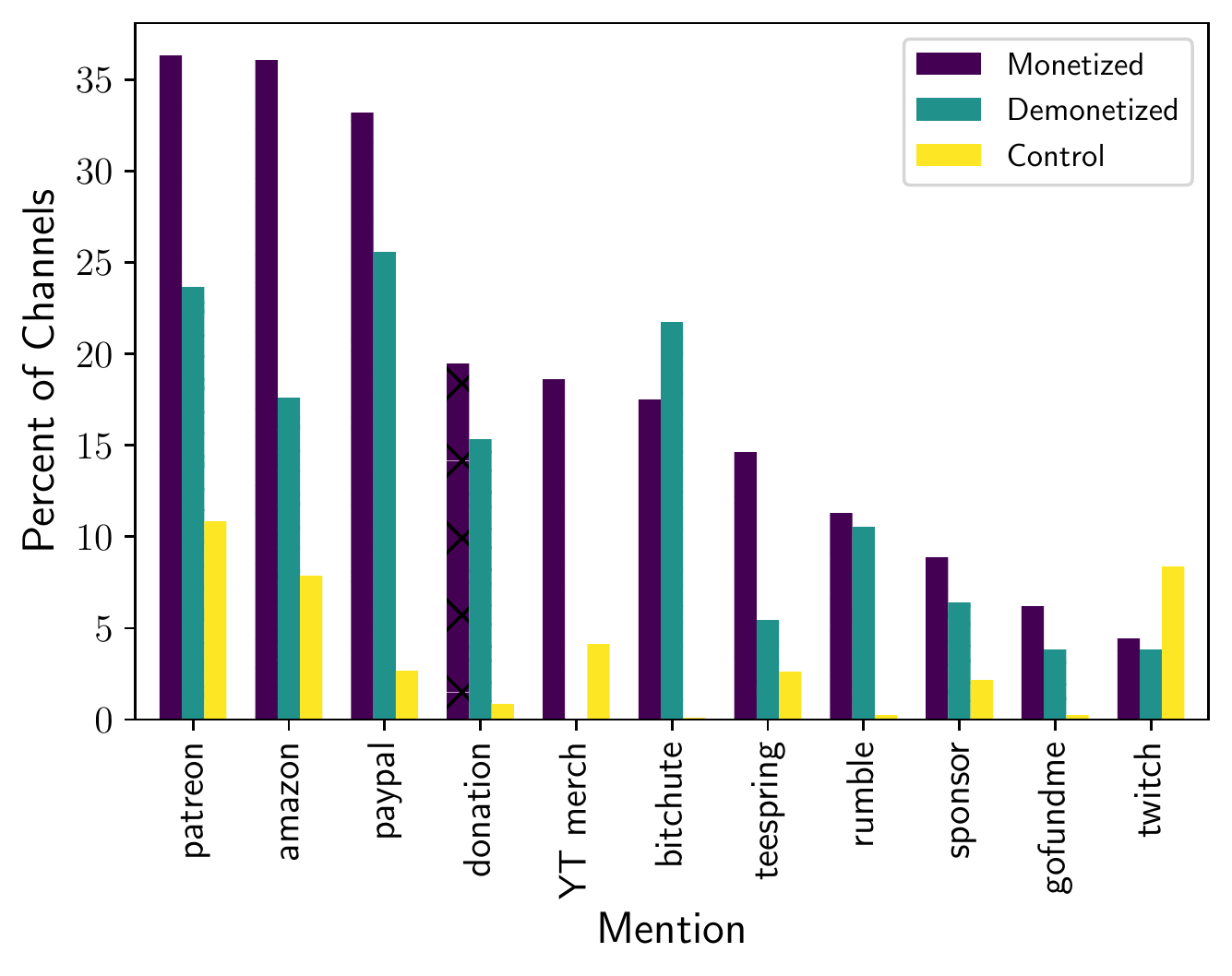}
    \caption{Percent of channels using offsite monetization methods. All except Twitch were more prevalent in conspiracy channels. ``YT merch'' is YouTube's official option for hosting offsite merchandise. All others are keywords found in video descriptions.}
    \label{fig:offsite_money}
\end{figure}

Even if they are demonetized, channels can still generate revenue by selling merchandise, asking for direct monetary support, or moving to alternative video hosting platforms. To understand the surrounding ecosystem of monetization used by creators, we analyzed links and keywords in video descriptions and flagged officially referenced merchandising pages. We identified the most popular offsite platforms by extracting URLs from video descriptions. Some URLs were shortened (e.g., with tinyurl), so we looked for the presence of specific keywords to identify references to offsite platforms. The additional keywords, ``donation'' and ``sponsor,'' were used to capture various other sites and calls to action linking to many different URLs, such as a channel's website. We split our analysis into monetized and likely demonetized conspiracy channels as well as channels from the control set.
Figure~\ref{fig:offsite_money} shows the results.

As a whole, conspiracy videos were 2.3 times as likely as control to use some form of offsite monetization ($p < 0.001$), and 5.2 times as likely to use multiple methods ($p < 0.001$). Patreon and Amazon were by far the most popular methods for monetization, with roughly 35\% of monetized conspiracy  channels linking to them. Excluding Twitch, at 8\% of control channels, all forms of monetization were more common in the conspiracy set. Alternative video hosting sites were also frequently linked to by conspiracy creators, with at least 17\% of conspiracy channels using Bitchute, and 11\% using Rumble. Notably, demonetized channels seemed to be disallowed from officially displaying merchandise, but circumvented the ban by linking to stores such as Teespring or Amazon.
Demonetized conspiracy channels were less likely to use all forms of offsite monetization, except for the alternative hosting site Bitchute (22\%). This suggests that explicit demonetization on YouTube is not the main reason for creators of conspiracy content to seek out alternative revenue, or alternatively, that creators of content demonetized by YouTube may find it difficult to monetize that content off-site, too. Supporting the latter conclusion, alternative platforms such as Patreon, PayPal, and Amazon banned QAnon conspiracy theorists and merchandise~\cite{qanonpatreon,qanonpaypal}. This could also explain why Bitchute and Rumble were comparatively more popular among demonetized channels. Both sites allow creators to directly monetize their content, and style themselves as ``free-speech'' platforms. %, making it an attractive option for creators that have been forced off other more mainstream sites.
Despite PayPal's removal of some QAnon accounts, it was the most popular option for demonetized channels, with 26\% mentioning the site.

Alternative social networks such as Parler, Telegram, and Gab were more common in the conspiracy set. Discord was more common in control. However, since these social networks are not often directly used for monetization, we excluded them from this analysis. Facebook and Twitter were also more prevalent in the conspiracy set, with Facebook nearly twice as common, but often links to these sites were in reference to specific posts or topics rather than to promote personal pages. 

It is worth noting that our results are most likely a lower bound on actual rates of offsite monetization. Many channels linked to individual websites that may contain products or donation links and would not be reflected in our results, and we did not analyze the actual video for offsite references either.

\subsection{Types of Advertisements}
A large majority of ad impressions were due to ads observed in both the conspiracy and the control data set.
The corresponding 1,405 unique advertisers (22\%) accounted for 90\% of control set impressions and 77\% of conspiracy, or a combined total of 160,084 ad impressions (87\%).
The two most frequent advertisers, totaladblock and Amazon's pharmacy, were each seen more than 20,000 times between both data sets. They presumably carry out minimal targeting and advertise all over YouTube, which makes these large advertisers less relevant for our study, except to note that their ads also appeared on conspiracy videos. The smaller rates of blanket advertising in the conspiracy set indicate more selective ad delivery.

\begin{table}[h!]
\centering
\footnotesize
\begin{tabular}{lSSr}
\toprule
Business Type           &  \hspace{-1em} Conspiracy &  Control & Residual $\triangledown$\\
\midrule
Self Improvement        &            20.6\,\% &         2.09\,\% &          104 \\
Special Interest Groups &            2.03\,\% &         1.78\,\% &         2.62 \\
Information Media       &            0.60\,\% &         0.47\,\% &         2.58 \\
Merchant                &            75.7\,\% &         84.0\,\% &        -30.6 \\
Aggregator Sites        &            1.04\,\% &         11.7\,\% &        -53.3 \\
\midrule
Content Type                   &                 &              &              \\
\midrule
Business                 &            11.6\,\% &         1.72\,\% &         69.3 \\
Alternative Health       &            7.34\,\% &         0.69\,\% &         61.4 \\
Lifestyle                &            5.49\,\% &         0.78\,\% &         47.5 \\
Electronics              &            5.82\,\% &         1.69\,\% &         35.7 \\
Education                &            4.44\,\% &         1.58\,\% &         26.9 \\
Food and Drink               &            6.37\,\% &         3.23\,\% &         22.6 \\
Entertainment            &            6.06\,\% &         3.38\,\% &         19.2 \\
Gold and Precious Metals &            0.54\,\% &         0.14\,\% &         11.4 \\
Insurance                &            1.30\,\% &         0.85\,\% &         6.54 \\
Beauty                   &            2.76\,\% &         2.18\,\% &         5.51 \\
Financial                &            5.84\,\% &         5.12\,\% &         4.54 \\
Games                    &            12.7\,\% &         11.9\,\% &         3.60 \\
Industrial               &            1.20\,\% &         0.96\,\% &         3.43 \\
Transportation           &            0.73\,\% &         0.93\,\% &        -3.14 \\
Government               &            0.06\,\% &         0.21\,\% &        -5.24 \\
Political                &            0.46\,\% &         0.84\,\% &        -6.33 \\
Home Goods               &            2.40\,\% &         4.67\,\% &        -16.3 \\
Fashion                  &            3.32\,\% &         6.66\,\% &        -20.3 \\
Major Retailer           &            3.73\,\% &         8.94\,\% &        -27.9 \\
Medical                  &            7.54\,\% &         16.2\,\% &        -35.5 \\
Software                 &            10.3\,\% &         27.3\,\% &        -57.5 \\
\bottomrule
\end{tabular}
\caption{Frequencies of advertising across business and content type. Columns \textit{Conspiracy} and \textit{Control} show each data set's percent of ad impressions from a particular category. \textit{Residual} shows the standard residual between the data sets. Positive values mean a category was more present in the conspiracy set. Any value larger than 3 is significant.}
\label{tab:catsubcat:results}
\end{table}

Despite the prevalence of common advertisers in both data sets, we found a significant difference in certain \textit{Business Types} between the ads seen on conspiracy and control videos ($\chi^2 = 13719.70$, $p < 0.001$, $df=4$). As seen in Table~\ref{tab:catsubcat:results}, self-improvement ads were much more likely to be shown on conspiracy videos. Many of these ads were from various webinars, books, or courses promising an easy route to financial independence; essentially digital advertising get-rich-quick schemes. These ads' presence is reflected in the greatly increased frequency of the Business content type.

Differences in the distribution of ads across \textit{Content Types} were also significant ($\chi^2 =17783.51$, $p < 0.001$, $df=22$). Notably, lifestyle and alternative health ads were much more common in the conspiracy data set than in the control data set. Insurance ads were also more common in the conspiracy set. This disparity was largely driven by two advertisers unique to the conspiracy set trying to 
generate leads for other insurance scammers and eventually sell subpar or entirely fake coverage. This type of predatory advertisement was outlined by a ProPublica report, warned about by the AARP, and the subject of at least one FTC lawsuit \cite{insurancescamsaarp,insurancescamspp}.

The prevalence of the Beauty, Electronics, and Entertainment content types in the conspiracy set was influenced by the presence of numerous deceptive affiliate marketing sites advertising assorted low-quality gadgets and beauty products. %, discussed further below.
Specific combinations of business and content type also appeared in greatly different frequencies. The subset of ads categorized as both self-improvement and business and finance was greatly overrepresented in the conspiracy set, with 9\% of all impressions in comparison to 0.2\% of impressions in control. Many of these ads made unreasonable promises of financial success. %, discussed in more detail in Section~\ref{sec:financialads}
% Self-improvement lifestyle ads also appeared at disproportionate rates in the conspiracy set, at 2.5\% vs.\ 0.2\%.
In conclusion, viewers of conspiracy videos appeared to be exposed to potentially predatory or deceptive ads at a much higher rate than viewers of mainstream content. We explore these types of ads in more detail in the next paragraphs.

\subsubsection{Predatory and Deceptive Marketing}
\label{sec:badads}
During the labeling process, we encountered various predatory and deceptive marketing practices. 
To estimate the prevalence of these practices in the full data sets, we identified common patterns in the URLs and web pages of predatory ads, such as affiliate IDs or tracking parameters used by specific companies. After parsing all URLs in the data sets to find all such ads, we spot-checked the results to verify correctness. %Since these pages are deceptive, they are hard to identify and
We erred on the side of caution to minimize %the number of
false positives. 

Broadly defined, predatory advertisers take advantage of consumer vulnerabilities to manipulate them into unfavorable market transactions. The specific characteristics that make a consumer ``vulnerable'' range from demographic classifications to informational asymmetries \cite{predads}. In the context of conspiracy theories, advertisers might be able to exploit an audience that is already more gullible, anxious about perceived threats from covert organizations, and mistrusting of government and regulatory bodies.

%\cbnote{control - 118 pred advertisers, 1926 impressions. Consp - 400 pred advertisers, 6462 impressions}
While we encountered deceptive styles of advertising in both data sets, they were disproportionately represented in the conspiracy set. Ads that we identified as deceptive or predatory accounted for 15\% of all impressions in the conspiracy set, but only 1.4\% of control impressions. Ad content ranged from get-rich-quick schemes, to promises of immortality through essential oils, to 5G-proof underwear. Admittedly, the content of many of these ads aligns with the content of the videos. Common themes of science denial and distrust of authority would naturally attract viewers supportive of natural medicine and tools to evade perceived governmental control. However, the selective ad delivery we observed allows advertisers of suspect products---such as the book that promises immortality seen on \textit{biblical-secret[.com]}---to reach audiences that might be susceptible to these scams while evading mainstream attention that un-targeted advertising could bring.

% \subsubsection{Financial Ads}
% \label{sec:financialads}
\noindent\textbf{Financial Ads.}
One of the more obviously predatory forms of advertising in the conspiracy set was the self-help business and finance ads.
The most common were webinars delivered by someone claiming to have achieved a high passive income through digital business, often explaining different affiliate marketing strategies, and similar to the ads in a recent FTC lawsuit~\cite{FTCscam}. These ``business'' courses accounted for 11\% of conspiracy, and less than 0.5\% of control ad impressions in the full data sets. While get-rich-quick scams are nothing new, their presence alongside dubious insurance ads and investment tips promising unreasonable returns indicates a pattern of predatory financial advertising on conspiracy videos. Given that automated ad targeting systems are designed to maximize clicks and successful sales, YouTube's advertising platform is potentially assisting financial scammers to find victims.

% \subsubsection{Dark Patterns and Deceptive Advertising Pages}
% \label{sec:decmarketing}

%\vznote{Method for looking for them here. Noticed affiliates and webinar often have similar parameters in query strings and patterns wrt outgoing links on their landing page. Used semiautomatic methods to check for these patterns, and spot checked them for accuracy. This doesn't catch all the affiliates/webinars, so results found here is probably just a fraction of what is actually happening.}

\noindent\textbf{Dark Patterns and Deceptive Advertising Pages}
Another common type of deceptive advertising used intermediary websites to promote many different low-quality products. These ads were almost non-existent in the control set, accounting for less than 0.1\% of ad impressions, compared to more than 3\% of conspiracy ad impressions. Advertisers hosted a site similar to that of the advertised product, but often with more elements of ``clickbait'' advertising, such as sensational claims and exciting images. The products advertised ranged from relatively harmless---such as an expensive ear cleaning device from \textit{buy-tvidler[.com]}---to actively dangerous, such as untested supplements claiming to cure Type-2 diabetes: \textit{buybloodsugarformula[.com]}. 

This style of deceptive advertising was supported extensively through affiliate marketing. Almost 95\% of the deceptive pages we found used some form of affiliate marketing, identified by affiliate IDs in the advertising URL or on the ad's web page. 
% By running many separate affiliate marketing campaigns and obscuring company ownership, these deceptive products are able to confuse consumers and potentially mitigate enforcement action against them.
Our estimate of deceptive affiliate activity is a lower bound; the sites are designed to deceive consumers and we erred on the side of caution for classification. In actuality, not all affiliate marketing is deceptive. Many Amazon ads in our data sets were affiliates advertising individual products. However, the vastly different proportions of deceptive marketing we encountered (more than 50 times more frequent in conspiracy) suggest an ad targeting system that enables selective delivery of low quality products (sometimes dangerously so) to specific audiences---in this case viewers of conspiracy theories. 

\section{Discussion}
\label{sec:discussion}

The significant difference in observed advertisements between control and conspiracy videos suggests that Google's ad delivery system enables, directly or indirectly, the targeting of conspiracy theory content.
Unfortunately, this resulted in a tenfold higher rate of deceptive and predatory advertising practices in our conspiracy set.
Apart from these advertisements, creators used third-party channels in an attempt to generate revenue from their content, even when demonetized by YouTube. All in all, our observations outline a robust network of monetization around conspiracy theories on YouTube, much of it at the expense of the audience.

\subsection{Recommendations}
Our research finds evidence for the success and necessity of cross-platform moderation efforts. While deplatforming content does reduce some of its reach~\cite{deplatforming}, demonetization alone does not prevent creators from using their audience on YouTube to leverage third-party monetization methods. We do find some evidence that creators who were demonetized on YouTube may have also been removed from other platforms, effectively reducing their chances at revenue generation. A more comprehensive understanding of the funding ecosystems for publishers of conspiracies and more general misinformation is necessary to address any financial incentives that encourage the proliferation of this content across the web. 

We also recommend that Google provide more transparent information about the ads it runs and how they are targeted, as well as content moderation outcomes such as demonetization.
Without further transparency into ads, we cannot tell whether the significant difference in advertising observed between mainstream and conspiracy YouTube videos is because of specific targeting performed by advertisers, or a more fundamental effect of YouTube's ad delivery optimization algorithm.
A wholistic understanding of online advertisers and the targeting systems they use would also enable greater consumer protections against the various scams enabled by digital advertising.
Without deeper visibility into the demonetization of individual videos or entire channels, we cannot distinguish with certainty whether an absence of advertising on a video is a choice of the video's creator, due to a lack of bids from advertisers, or because of a ban on advertising imposed by YouTube.
Understanding the level of monetization is also important for other types of problematic content that we did not explore in this paper, such as hateful and extremist content.

\subsection{Limitations and Future Work}
\label{sec:limitations}

The opacity and complexity of Google's ad targeting system renders any definitive conclusions difficult to make. Ads are served based on a combination of targeting information---behavioral or content-based---and decisions on delivery---optimizing for clicks or sales---within given campaign parameters. Restricting ad personalization allows us to make conclusions about content-based targeting, but may prevent us from seeing ads only served through behavioral targeting. Furthermore, outside of any targeting by the advertiser, the delivery system of Google ads may influence or wholly cause the difference we see between the conspiracy and control set; ads with lower bids (cost per click), for example, could be shown more often to less ``valuable'' traffic.
Lastly, we do not know whether the lower advertisement rate of conspiracy videos is voluntary or involuntary, i.e., whether content creators are actually seeking to make a profit.
Some of the differences we observed may also be due to our choice of control set with different types, age and popularity of videos, and different audience demographics compared to conspiracy videos, e.g.\ less popular videos could be more likely to contain scam ads, regardless of being conspiratorial.
When investigating off-platform monetization strategies of conspiracy content, we did not attempt to infer the intent behind linking to PayPal or Patreon (i.e., whether creators are seeking donations for themselves or for a third party), or when linking to another social media site (i.e., whether promoting their own content on a different platform, or referring to a third-party post). 
It is clear someone is profiting from these channels, but more research is necessary to understand the full network of off-site monetization strategies, and the degree to which channels can generate revenue from them.

\section{Conclusion}
Our study has characterized the methods and types of monetization on conspiracy theory YouTube videos. We find a significantly higher prevalence of predatory advertisements on conspiracy videos, as well as many third-party alternative revenue generation opportunities. Assuming that Google's ad delivery system successfully optimizes for clicks and successful sales, the difference in advertising quality suggests that YouTube's advertising platform may be assisting predatory advertisers to identify potential victims. While our study is a significant first exploration of the monetization of conspiracy theories, more research is necessary to fully characterize and understand the broader financial ecosystem for publishers of false information and the advertisers that profit off of them. Increased transparency around advertising and moderation practices from major platforms such as Google would go a long way towards improving research and knowledge in this area.

%%
%% The acknowledgments section is defined using the "acks" environment
%% (and NOT an unnumbered section). This ensures the proper
%% identification of the section in the article metadata, and the
%% consistent spelling of the heading.
\begin{acks}
Cybersecurity for Democracy at NYU’s Center for Cybersecurity has been supported by \grantsponsor{1}{Democracy Fund}{https://democracyfund.org/}, \grantsponsor{2}{Luminate}{https://luminategroup.com}, \grantsponsor{3}{Media Democracy Fund}{https://mediademocracyfund.org/}, the \grantsponsor{4}{National Science Foundation}{https://www.nsf.gov/} under grant \grantnum{4}{1814816}, % https://www.nsf.gov/awardsearch/showAward?AWD_ID=1814816&amp\%3BHistoricalAwards=false
\grantsponsor{5}{Reset}{https://luminategroup.com/reset}, and \grantsponsor{6}{Wellspring}{https://wpfund.org/}. The Privacy, Security, and Automation Lab at NYU's CCS has been supported by \grantsponsor{4}{NSF}{https://www.nsf.gov/} grant \grantnum{4}{1931005}. % [https://www.nsf.gov/awardsearch/showAward?AWD_ID=1931005&HistoricalAwards=false]
Our labs have also received gifts from Google. Any opinions, findings, and conclusions or recommendations expressed in this paper are those of the authors and do not necessarily reflect the view of our funders.
\end{acks}

%%
%% The next two lines define the bibliography style to be used, and
%% the bibliography file.
\bibliographystyle{ACM-Reference-Format}
\bibliography{references}

%%% -*-BibTeX-*-
%%% Do NOT edit. File created by BibTeX with style
%%% ACM-Reference-Format-Journals [18-Jan-2012].

\begin{thebibliography}{70}

%%% ====================================================================
%%% NOTE TO THE USER: you can override these defaults by providing
%%% customized versions of any of these macros before the \bibliography
%%% command.  Each of them MUST provide its own final punctuation,
%%% except for \shownote{}, \showDOI{}, and \showURL{}.  The latter two
%%% do not use final punctuation, in order to avoid confusing it with
%%% the Web address.
%%%
%%% To suppress output of a particular field, define its macro to expand
%%% to an empty string, or better, \unskip, like this:
%%%
%%% \newcommand{\showDOI}[1]{\unskip}   % LaTeX syntax
%%%
%%% \def \showDOI #1{\unskip}           % plain TeX syntax
%%%
%%% ====================================================================

\ifx \showCODEN    \undefined \def \showCODEN     #1{\unskip}     \fi
\ifx \showDOI      \undefined \def \showDOI       #1{#1}\fi
\ifx \showISBNx    \undefined \def \showISBNx     #1{\unskip}     \fi
\ifx \showISBNxiii \undefined \def \showISBNxiii  #1{\unskip}     \fi
\ifx \showISSN     \undefined \def \showISSN      #1{\unskip}     \fi
\ifx \showLCCN     \undefined \def \showLCCN      #1{\unskip}     \fi
\ifx \shownote     \undefined \def \shownote      #1{#1}          \fi
\ifx \showarticletitle \undefined \def \showarticletitle #1{#1}   \fi
\ifx \showURL      \undefined \def \showURL       {\relax}        \fi
% The following commands are used for tagged output and should be
% invisible to TeX
\providecommand\bibfield[2]{#2}
\providecommand\bibinfo[2]{#2}
\providecommand\natexlab[1]{#1}
\providecommand\showeprint[2][]{arXiv:#2}

\bibitem[\protect\citeauthoryear{AARP}{AARP}{2020}]%
        {insurancescamsaarp}
\bibfield{author}{\bibinfo{person}{AARP}.} \bibinfo{year}{2020}\natexlab{}.
\newblock \bibinfo{booktitle}{\emph{Health Insurance Scams}}.
\newblock
\urldef\tempurl%
\url{https://www.aarp.org/money/scams-fraud/info-2019/health-insurance.html}
\showURL{%
\tempurl}


\bibitem[\protect\citeauthoryear{Alba}{Alba}{2017}]%
        {alba17}
\bibfield{author}{\bibinfo{person}{Davey Alba}.}
  \bibinfo{year}{2017}\natexlab{}.
\newblock \bibinfo{booktitle}{\emph{The Best Way to Quash Fake News? Choke Off
  Its Ad Money}}.
\newblock
\urldef\tempurl%
\url{https://www.wired.com/2017/02/best-way-quash-fake-news-choke-off-ad-money/}
\showURL{%
\tempurl}


\bibitem[\protect\citeauthoryear{Ali, Sapiezynski, Bogen, Korolova, Mislove,
  and Rieke}{Ali et~al\mbox{.}}{2019}]%
        {ali19}
\bibfield{author}{\bibinfo{person}{Muhammad Ali}, \bibinfo{person}{Piotr
  Sapiezynski}, \bibinfo{person}{Miranda Bogen}, \bibinfo{person}{Aleksandra
  Korolova}, \bibinfo{person}{Alan Mislove}, {and} \bibinfo{person}{Aaron
  Rieke}.} \bibinfo{year}{2019}\natexlab{}.
\newblock \showarticletitle{Discrimination through Optimization: How Facebook's
  Ad Delivery Can Lead to Biased Outcomes}.
\newblock \bibinfo{journal}{\emph{Proc. ACM Hum.-Comput. Interact.}}
  \bibinfo{volume}{3}, \bibinfo{number}{CSCW}, Article \bibinfo{articleno}{199}
  (\bibinfo{date}{Nov.} \bibinfo{year}{2019}), \bibinfo{numpages}{30}~pages.
\newblock
\urldef\tempurl%
\url{https://doi.org/10.1145/3359301}
\showDOI{\tempurl}


\bibitem[\protect\citeauthoryear{Ali, Saeed, Aldreabi, Blackburn,
  De~Cristofaro, Zannettou, and Stringhini}{Ali et~al\mbox{.}}{2021}]%
        {deplatforming}
\bibfield{author}{\bibinfo{person}{Shiza Ali}, \bibinfo{person}{Mohammad~Hammas
  Saeed}, \bibinfo{person}{Esraa Aldreabi}, \bibinfo{person}{Jeremy Blackburn},
  \bibinfo{person}{Emiliano De~Cristofaro}, \bibinfo{person}{Savvas Zannettou},
  {and} \bibinfo{person}{Gianluca Stringhini}.}
  \bibinfo{year}{2021}\natexlab{}.
\newblock \showarticletitle{Understanding the Effect of Deplatforming on Social
  Networks}. In \bibinfo{booktitle}{\emph{13th ACM Web Science Conference
  2021}} (Virtual Event, United Kingdom) \emph{(\bibinfo{series}{WebSci '21})}.
  \bibinfo{publisher}{Association for Computing Machinery},
  \bibinfo{address}{New York, NY, USA}, \bibinfo{pages}{187–195}.
\newblock
\showISBNx{9781450383301}
\urldef\tempurl%
\url{https://doi.org/10.1145/3447535.3462637}
\showDOI{\tempurl}


\bibitem[\protect\citeauthoryear{Aliapoulios, Papasavva, Ballard, Cristofaro,
  Stringhini, Zannettou, and Blackburn}{Aliapoulios et~al\mbox{.}}{2021}]%
        {aliapoulios21}
\bibfield{author}{\bibinfo{person}{Max Aliapoulios}, \bibinfo{person}{Antonis
  Papasavva}, \bibinfo{person}{Cameron Ballard}, \bibinfo{person}{Emiliano~De
  Cristofaro}, \bibinfo{person}{Gianluca Stringhini}, \bibinfo{person}{Savvas
  Zannettou}, {and} \bibinfo{person}{Jeremy Blackburn}.}
  \bibinfo{year}{2021}\natexlab{}.
\newblock \bibinfo{title}{The Gospel According to Q: Understanding the QAnon
  Conspiracy from the Perspective of Canonical Information}.
\newblock
\newblock
\showeprint[arxiv]{2101.08750}~[cs.CY]


\bibitem[\protect\citeauthoryear{Allcott, Gentzkow, and Yu}{Allcott
  et~al\mbox{.}}{2019}]%
        {Allcott19}
\bibfield{author}{\bibinfo{person}{Hunt Allcott}, \bibinfo{person}{Matthew
  Gentzkow}, {and} \bibinfo{person}{Chuan Yu}.}
  \bibinfo{year}{2019}\natexlab{}.
\newblock \showarticletitle{Trends in the diffusion of misinformation on social
  media}.
\newblock \bibinfo{journal}{\emph{Research \& Politics}} \bibinfo{volume}{6},
  \bibinfo{number}{2} (\bibinfo{year}{2019}),
  \bibinfo{pages}{2053168019848554}.
\newblock
\urldef\tempurl%
\url{https://doi.org/10.1177/2053168019848554}
\showDOI{\tempurl}
\showeprint{https://doi.org/10.1177/2053168019848554}


\bibitem[\protect\citeauthoryear{Asmolov}{Asmolov}{2018}]%
        {Asmolov18}
\bibfield{author}{\bibinfo{person}{Gregory Asmolov}.}
  \bibinfo{year}{2018}\natexlab{}.
\newblock \showarticletitle{The disconnective power of disinformation
  campaigns}.
\newblock \bibinfo{journal}{\emph{Journal of International Affairs}}
  \bibinfo{volume}{71}, \bibinfo{number}{1.5} (\bibinfo{year}{2018}),
  \bibinfo{pages}{69--76}.
\newblock
\showISSN{0022197X}
\urldef\tempurl%
\url{https://www.jstor.org/stable/26508120}
\showURL{%
\tempurl}


\bibitem[\protect\citeauthoryear{Barford, Canadi, Krushevskaja, Ma, and
  Muthukrishnan}{Barford et~al\mbox{.}}{2014}]%
        {barford14}
\bibfield{author}{\bibinfo{person}{Paul Barford}, \bibinfo{person}{Igor
  Canadi}, \bibinfo{person}{Darja Krushevskaja}, \bibinfo{person}{Qiang Ma},
  {and} \bibinfo{person}{S. Muthukrishnan}.} \bibinfo{year}{2014}\natexlab{}.
\newblock \showarticletitle{Adscape: Harvesting and Analyzing Online Display
  Ads}. In \bibinfo{booktitle}{\emph{Proceedings of the 23rd International
  Conference on World Wide Web}} (Seoul, Korea) \emph{(\bibinfo{series}{WWW
  '14})}. \bibinfo{publisher}{Association for Computing Machinery},
  \bibinfo{address}{New York, NY, USA}, \bibinfo{pages}{597–608}.
\newblock
\showISBNx{9781450327442}
\urldef\tempurl%
\url{https://doi.org/10.1145/2566486.2567992}
\showDOI{\tempurl}


\bibitem[\protect\citeauthoryear{Bleakley}{Bleakley}{2021}]%
        {qanonrise}
\bibfield{author}{\bibinfo{person}{Paul Bleakley}.}
  \bibinfo{year}{2021}\natexlab{}.
\newblock \showarticletitle{Panic, pizza and mainstreaming the alt-right: A
  social media analysis of Pizzagate and the rise of the QAnon conspiracy}.
\newblock \bibinfo{journal}{\emph{Current Sociology}} (\bibinfo{date}{29 July}
  \bibinfo{year}{2021}).
\newblock
\urldef\tempurl%
\url{https://doi.org/10.1177/00113921211034896}
\showURL{%
\tempurl}


\bibitem[\protect\citeauthoryear{Bradshaw and Howard}{Bradshaw and
  Howard}{2018}]%
        {Bradshaw18}
\bibfield{author}{\bibinfo{person}{Samantha Bradshaw} {and}
  \bibinfo{person}{Philip~N. Howard}.} \bibinfo{year}{2018}\natexlab{}.
\newblock \showarticletitle{The global organization of social media
  disinformation campaigns}.
\newblock \bibinfo{journal}{\emph{Journal of International Affairs}}
  \bibinfo{volume}{71}, \bibinfo{number}{1.5} (\bibinfo{year}{2018}),
  \bibinfo{pages}{23--32}.
\newblock
\showISSN{0022197X}
\urldef\tempurl%
\url{https://www.jstor.org/stable/26508115}
\showURL{%
\tempurl}


\bibitem[\protect\citeauthoryear{Braun and Eklund}{Braun and Eklund}{2019}]%
        {braun19}
\bibfield{author}{\bibinfo{person}{Joshua~A. Braun} {and}
  \bibinfo{person}{Jessica~L. Eklund}.} \bibinfo{year}{2019}\natexlab{}.
\newblock \showarticletitle{Fake News, Real Money: Ad Tech Platforms,
  Profit-Driven Hoaxes, and the Business of Journalism}.
\newblock \bibinfo{journal}{\emph{Digital Journalism}} \bibinfo{volume}{7},
  \bibinfo{number}{1} (\bibinfo{year}{2019}), \bibinfo{pages}{1--21}.
\newblock
\urldef\tempurl%
\url{https://doi.org/10.1080/21670811.2018.1556314}
\showDOI{\tempurl}
\showeprint{https://doi.org/10.1080/21670811.2018.1556314}


\bibitem[\protect\citeauthoryear{Brennen, Simon, Howard, and Nielsen}{Brennen
  et~al\mbox{.}}{2020}]%
        {reuters20}
\bibfield{author}{\bibinfo{person}{J~Scott Brennen}, \bibinfo{person}{Felix
  Simon}, \bibinfo{person}{Phillip~N Howard}, {and}
  \bibinfo{person}{Rasmus~Klein Nielsen}.} \bibinfo{year}{2020}\natexlab{}.
\newblock \showarticletitle{Types, sources, and claims of COVID-19
  misinformation}.
\newblock \bibinfo{journal}{\emph{Reuters Institute}} (\bibinfo{year}{2020}).
\newblock
\urldef\tempurl%
\url{https://reutersinstitute.politics.ox.ac.uk/types-sources-and-claims-covid-19-misinformation}
\showURL{%
\tempurl}


\bibitem[\protect\citeauthoryear{Clark and Zaitsev}{Clark and Zaitsev}{2020}]%
        {chan2vec}
\bibfield{author}{\bibinfo{person}{Sam Clark} {and} \bibinfo{person}{Anna
  Zaitsev}.} \bibinfo{year}{2020}\natexlab{}.
\newblock \bibinfo{title}{Understanding YouTube Communities via
  Subscription-based Channel Embeddings}.
\newblock
\newblock
\showeprint[arxiv]{2010.09892}~[cs.LG]


\bibitem[\protect\citeauthoryear{Commission}{Commission}{2021}]%
        {FTCscam}
\bibfield{author}{\bibinfo{person}{Federal~Trade Commission}.}
  \bibinfo{year}{2021}\natexlab{}.
\newblock \bibinfo{booktitle}{\emph{FTC Returns \$1.1 Million to Consumers Who
  Lost Money to Alleged Scammers Selling Bogus Income Opportunities}}.
\newblock
\urldef\tempurl%
\url{https://www.ftc.gov/news-events/press-releases/2021/10/ftc-returns-11-million-consumers-who-lost-money-alleged-scammers}
\showURL{%
\tempurl}


\bibitem[\protect\citeauthoryear{Dastin, Dang, and Irrera}{Dastin
  et~al\mbox{.}}{2021}]%
        {qanonpaypal}
\bibfield{author}{\bibinfo{person}{Jeffrey Dastin}, \bibinfo{person}{Sheila
  Dang}, {and} \bibinfo{person}{Anna Irrera}.} \bibinfo{year}{2021}\natexlab{}.
\newblock \bibinfo{booktitle}{\emph{Online merchants linked to QAnon down, but
  not out, following platform bans}}.
\newblock
\urldef\tempurl%
\url{https://www.reuters.com/article/us-usa-trump-qanon-financing/online-merchants-linked-to-qanon-down-but-not-out-following-platform-bans-idUSKBN29U193}
\showURL{%
\tempurl}


\bibitem[\protect\citeauthoryear{Datta, Datta, Makagon, Mulligan, and
  Tschantz}{Datta et~al\mbox{.}}{2018}]%
        {datta18}
\bibfield{author}{\bibinfo{person}{Amit Datta}, \bibinfo{person}{Anupam Datta},
  \bibinfo{person}{Jael Makagon}, \bibinfo{person}{Deirdre~K. Mulligan}, {and}
  \bibinfo{person}{Michael~Carl Tschantz}.} \bibinfo{year}{2018}\natexlab{}.
\newblock \showarticletitle{Discrimination in Online Advertising: A
  Multidisciplinary Inquiry}. \bibinfo{publisher}{Proceedings of Machine
  Learning Research}.
\newblock


\bibitem[\protect\citeauthoryear{Datta, Tschantz, and Datta}{Datta
  et~al\mbox{.}}{2014}]%
        {adfisher}
\bibfield{author}{\bibinfo{person}{Amit Datta}, \bibinfo{person}{Michael~Carl
  Tschantz}, {and} \bibinfo{person}{Anupam Datta}.}
  \bibinfo{year}{2014}\natexlab{}.
\newblock \showarticletitle{Automated Experiments on Ad Privacy Settings: A
  Tale of Opacity, Choice, and Discrimination}.
\newblock \bibinfo{journal}{\emph{ArXiv}}  \bibinfo{volume}{abs/1408.6491}
  (\bibinfo{year}{2014}).
\newblock


\bibitem[\protect\citeauthoryear{de~Zeeuw, Hagen, Peeters, and
  Jokubauskaite}{de~Zeeuw et~al\mbox{.}}{2020}]%
        {Normiefication}
\bibfield{author}{\bibinfo{person}{Daniel de Zeeuw}, \bibinfo{person}{Sal
  Hagen}, \bibinfo{person}{Stijn Peeters}, {and} \bibinfo{person}{Emilija
  Jokubauskaite}.} \bibinfo{year}{2020}\natexlab{}.
\newblock \showarticletitle{Tracing normiefication: A cross-platform analysis
  of the QAnon conspiracy theory}.
\newblock \bibinfo{journal}{\emph{First Monday}} \bibinfo{volume}{25},
  \bibinfo{number}{11} (\bibinfo{date}{Oct.} \bibinfo{year}{2020}).
\newblock
\urldef\tempurl%
\url{https://doi.org/10.5210/fm.v25i11.10643}
\showDOI{\tempurl}


\bibitem[\protect\citeauthoryear{Dehghani, Niaki, Ramezani, and Sali}{Dehghani
  et~al\mbox{.}}{2016}]%
        {dehghani16}
\bibfield{author}{\bibinfo{person}{Milad Dehghani},
  \bibinfo{person}{Mojtaba~Khorram Niaki}, \bibinfo{person}{Iman Ramezani},
  {and} \bibinfo{person}{Rasoul Sali}.} \bibinfo{year}{2016}\natexlab{}.
\newblock \showarticletitle{Evaluating the influence of YouTube advertising for
  attraction of young customers}.
\newblock \bibinfo{journal}{\emph{Computers in Human Behavior}}
  \bibinfo{volume}{59} (\bibinfo{year}{2016}), \bibinfo{pages}{165--172}.
\newblock
\showISSN{0747-5632}
\urldef\tempurl%
\url{https://doi.org/10.1016/j.chb.2016.01.037}
\showDOI{\tempurl}


\bibitem[\protect\citeauthoryear{Del~Vicario, Bessi, Zollo, Petroni, Scala,
  Caldarelli, Stanley, and Quattrociocchi}{Del~Vicario et~al\mbox{.}}{2016}]%
        {Vicario16}
\bibfield{author}{\bibinfo{person}{Michela Del~Vicario},
  \bibinfo{person}{Alessandro Bessi}, \bibinfo{person}{Fabiana Zollo},
  \bibinfo{person}{Fabio Petroni}, \bibinfo{person}{Antonio Scala},
  \bibinfo{person}{Guido Caldarelli}, \bibinfo{person}{H.~Eugene Stanley},
  {and} \bibinfo{person}{Walter Quattrociocchi}.}
  \bibinfo{year}{2016}\natexlab{}.
\newblock \showarticletitle{The spreading of misinformation online}.
\newblock \bibinfo{journal}{\emph{Proceedings of the National Academy of
  Sciences}} \bibinfo{volume}{113}, \bibinfo{number}{3} (\bibinfo{year}{2016}),
  \bibinfo{pages}{554--559}.
\newblock
\showISSN{0027-8424}
\urldef\tempurl%
\url{https://doi.org/10.1073/pnas.1517441113}
\showDOI{\tempurl}
\showeprint{https://www.pnas.org/content/113/3/554.full.pdf}


\bibitem[\protect\citeauthoryear{Douglas, Sutton, and Cichocka}{Douglas
  et~al\mbox{.}}{2019}]%
        {notgullible}
\bibfield{author}{\bibinfo{person}{Karen Douglas}, \bibinfo{person}{Robbie
  Sutton}, {and} \bibinfo{person}{Aleksandra Cichocka}.}
  \bibinfo{year}{2019}\natexlab{}.
\newblock \bibinfo{booktitle}{\emph{Belief in Conspiracy Theories: Looking
  beyond gullibility}}.
\newblock \bibinfo{pages}{61--76}.
\newblock
\showISBNx{9780429203787}
\urldef\tempurl%
\url{https://doi.org/10.4324/9780429203787-4}
\showDOI{\tempurl}


\bibitem[\protect\citeauthoryear{Edelson, Lauinger, and McCoy}{Edelson
  et~al\mbox{.}}{2020}]%
        {edelson20}
\bibfield{author}{\bibinfo{person}{Laura Edelson}, \bibinfo{person}{Tobias
  Lauinger}, {and} \bibinfo{person}{Damon McCoy}.}
  \bibinfo{year}{2020}\natexlab{}.
\newblock \showarticletitle{A Security Analysis of the Facebook Ad Library}. In
  \bibinfo{booktitle}{\emph{2020 IEEE Symposium on Security and Privacy (SP)}}.
  \bibinfo{pages}{661--678}.
\newblock
\urldef\tempurl%
\url{https://doi.org/10.1109/SP40000.2020.00084}
\showDOI{\tempurl}


\bibitem[\protect\citeauthoryear{Espinoza and Stefano}{Espinoza and
  Stefano}{2020}]%
        {espinoza20}
\bibfield{author}{\bibinfo{person}{Javier Espinoza} {and}
  \bibinfo{person}{Mark~Di Stefano}.} \bibinfo{year}{2020}\natexlab{}.
\newblock \bibinfo{booktitle}{\emph{Fake news websites still profit from Google
  advertising}}.
\newblock
\urldef\tempurl%
\url{https://www.ft.com/content/5f8a405c-c132-4d9b-a86f-c52884535f3e}
\showURL{%
\tempurl}


\bibitem[\protect\citeauthoryear{Faddoul, Chaslot, and Farid}{Faddoul
  et~al\mbox{.}}{2020}]%
        {faddoul20}
\bibfield{author}{\bibinfo{person}{Marc Faddoul}, \bibinfo{person}{Guillaume
  Chaslot}, {and} \bibinfo{person}{Hany Farid}.}
  \bibinfo{year}{2020}\natexlab{}.
\newblock \showarticletitle{A Longitudinal Analysis of YouTube's Promotion of
  Conspiracy Videos}.
\newblock \bibinfo{journal}{\emph{CoRR}}  \bibinfo{volume}{abs/2003.03318}
  (\bibinfo{year}{2020}).
\newblock
\showeprint[arXiv]{2003.03318}
\urldef\tempurl%
\url{https://arxiv.org/abs/2003.03318}
\showURL{%
\tempurl}


\bibitem[\protect\citeauthoryear{Farivar}{Farivar}{2021}]%
        {qanonriots}
\bibfield{author}{\bibinfo{person}{Masood Farivar}.}
  \bibinfo{year}{2021}\natexlab{}.
\newblock \bibinfo{booktitle}{\emph{Capitol Riot Exposed QAnon’s Violent
  Potential}}.
\newblock
\urldef\tempurl%
\url{https://www.voanews.com/a/usa_capitol-riot-exposed-qanons-violent-potential/6203967.html}
\showURL{%
\tempurl}


\bibitem[\protect\citeauthoryear{Funke, Benkelman, and Tardaguila}{Funke
  et~al\mbox{.}}{2019}]%
        {funke19}
\bibfield{author}{\bibinfo{person}{Daniel Funke}, \bibinfo{person}{Susan
  Benkelman}, {and} \bibinfo{person}{Cristina Tardaguila}.}
  \bibinfo{year}{2019}\natexlab{}.
\newblock \bibinfo{booktitle}{\emph{Factually: How misinformation makes
  money}}.
\newblock
\urldef\tempurl%
\url{https://www.americanpressinstitute.org/fact-checking-project/factually-newsletter/factually-how-misinformation-makes-money/}
\showURL{%
\tempurl}


\bibitem[\protect\citeauthoryear{GARRETT and TOUMANOFF}{GARRETT and
  TOUMANOFF}{2010}]%
        {predads}
\bibfield{author}{\bibinfo{person}{DENNIS~E. GARRETT} {and}
  \bibinfo{person}{PETER~G. TOUMANOFF}.} \bibinfo{year}{2010}\natexlab{}.
\newblock \showarticletitle{Are Consumers Disadvantaged or Vulnerable? An
  Examination of Consumer Complaints to the Better Business Bureau}.
\newblock \bibinfo{journal}{\emph{Journal of Consumer Affairs}}
  \bibinfo{volume}{44}, \bibinfo{number}{1} (\bibinfo{year}{2010}),
  \bibinfo{pages}{3--23}.
\newblock
\urldef\tempurl%
\url{https://doi.org/10.1111/j.1745-6606.2010.01155.x}
\showDOI{\tempurl}
\showeprint{https://onlinelibrary.wiley.com/doi/pdf/10.1111/j.1745-6606.2010.01155.x}


\bibitem[\protect\citeauthoryear{Garry, Walter, Rukaya, and Mohhamed}{Garry
  et~al\mbox{.}}{2021}]%
        {Qanonderad}
\bibfield{author}{\bibinfo{person}{Amanda Garry}, \bibinfo{person}{Samantha
  Walter}, \bibinfo{person}{Rukaya Rukaya}, {and} \bibinfo{person}{Ayan
  Mohhamed}.} \bibinfo{year}{2021}\natexlab{}.
\newblock \showarticletitle{QAnon Conspiracy Theory: Examining its Evolution
  and Mechanisms of Radicalization}.
\newblock \bibinfo{journal}{\emph{Journal for Deradicalization}}
  (\bibinfo{date}{25 Mar} \bibinfo{year}{2021}).
\newblock
\urldef\tempurl%
\url{https://journals.sfu.ca/jd/index.php/jd/article/view/437}
\showURL{%
\tempurl}


\bibitem[\protect\citeauthoryear{Google}{Google}{2021a}]%
        {Ytadfriendliness}
\bibfield{author}{\bibinfo{person}{Google}.} \bibinfo{year}{2021}\natexlab{a}.
\newblock \bibinfo{booktitle}{\emph{Advertiser-friendly content guidelines}}.
\newblock
\urldef\tempurl%
\url{https://support.google.com/youtube/answer/6162278}
\showURL{%
\tempurl}


\bibitem[\protect\citeauthoryear{Google}{Google}{2021b}]%
        {Ytguidelines}
\bibfield{author}{\bibinfo{person}{Google}.} \bibinfo{year}{2021}\natexlab{b}.
\newblock \bibinfo{booktitle}{\emph{YouTube channel monetization policies}}.
\newblock
\urldef\tempurl%
\url{https://support.google.com/youtube/answer/1311392}
\showURL{%
\tempurl}


\bibitem[\protect\citeauthoryear{Google}{Google}{2021c}]%
        {Ytpartner}
\bibfield{author}{\bibinfo{person}{Google}.} \bibinfo{year}{2021}\natexlab{c}.
\newblock \bibinfo{booktitle}{\emph{YouTube Partner Program overview \&
  eligibility}}.
\newblock
\urldef\tempurl%
\url{https://support.google.com/youtube/answer/72851?hl=en}
\showURL{%
\tempurl}


\bibitem[\protect\citeauthoryear{Graham}{Graham}{2020}]%
        {Policychange}
\bibfield{author}{\bibinfo{person}{Megan Graham}.}
  \bibinfo{year}{2020}\natexlab{}.
\newblock \bibinfo{booktitle}{}.
\newblock
\urldef\tempurl%
\url{https://www.cnbc.com/2020/11/19/youtube-will-put-ads-on-non-partner-videos-but-wont-pay-the-creators.html}
\showURL{%
\tempurl}


\bibitem[\protect\citeauthoryear{Greenspan}{Greenspan}{2020}]%
        {qanonpatreon}
\bibfield{author}{\bibinfo{person}{Rachel~E Greenspan}.}
  \bibinfo{year}{2020}\natexlab{}.
\newblock \bibinfo{booktitle}{\emph{Patreon is banning QAnon conspiracy
  theorists, joining a growing group of tech companies taking action against
  the movement}}.
\newblock
\urldef\tempurl%
\url{https://www.businessinsider.com/patreon-bans-qanon-conspiracy-theory-users-latest-tech-company-2020-10}
\showURL{%
\tempurl}


\bibitem[\protect\citeauthoryear{Herley}{Herley}{2012}]%
        {nigerianscammers}
\bibfield{author}{\bibinfo{person}{Cormac Herley}.}
  \bibinfo{year}{2012}\natexlab{}.
\newblock \showarticletitle{Why do Nigerian Scammers Say They are from
  Nigeria?}
\newblock \bibinfo{journal}{\emph{Proceedings of the Workshop on the Economics
  of Information Security}} (\bibinfo{date}{01} \bibinfo{year}{2012}).
\newblock


\bibitem[\protect\citeauthoryear{Hughes, Kojm, Lal, and Tromble}{Hughes
  et~al\mbox{.}}{2021}]%
        {qanonriots2}
\bibfield{author}{\bibinfo{person}{Seamus Hughes},
  \bibinfo{person}{Cristopher~A. Kojm}, \bibinfo{person}{Rollie Lal}, {and}
  \bibinfo{person}{Rebekah Tromble}.} \bibinfo{year}{2021}\natexlab{}.
\newblock \bibinfo{booktitle}{\emph{The Capitol Riots, QAnon, and the
  Internet}}.
\newblock
\urldef\tempurl%
\url{https://iddp.gwu.edu/capitol-riots-qanon-and-internet}
\showURL{%
\tempurl}


\bibitem[\protect\citeauthoryear{Hussein, Juneja, and Mitra}{Hussein
  et~al\mbox{.}}{2020}]%
        {hussein20}
\bibfield{author}{\bibinfo{person}{Eslam Hussein}, \bibinfo{person}{Prerna
  Juneja}, {and} \bibinfo{person}{Tanushree Mitra}.}
  \bibinfo{year}{2020}\natexlab{}.
\newblock \showarticletitle{Measuring Misinformation in Video Search Platforms:
  An Audit Study on YouTube}.
\newblock \bibinfo{journal}{\emph{Proceedings of the ACM on Human-Computer
  Interaction}}  \bibinfo{volume}{4} (\bibinfo{year}{2020}), \bibinfo{pages}{1
  -- 27}.
\newblock


\bibitem[\protect\citeauthoryear{Jagtap, Kumar, Goel, Sharma, Sharma, and
  George}{Jagtap et~al\mbox{.}}{2021}]%
        {jagtap21}
\bibfield{author}{\bibinfo{person}{Rajlaxmi Jagtap}, \bibinfo{person}{Abhinav
  Kumar}, \bibinfo{person}{Rahul Goel}, \bibinfo{person}{Shakshi Sharma},
  \bibinfo{person}{Rajesh Sharma}, {and} \bibinfo{person}{Clint~P. George}.}
  \bibinfo{year}{2021}\natexlab{}.
\newblock \showarticletitle{Misinformation Detection on YouTube Using Video
  Captions}.
\newblock \bibinfo{journal}{\emph{ArXiv}}  \bibinfo{volume}{abs/2107.00941}
  (\bibinfo{year}{2021}).
\newblock


\bibitem[\protect\citeauthoryear{Jolley and Douglas}{Jolley and
  Douglas}{2014}]%
        {jolley14}
\bibfield{author}{\bibinfo{person}{Daniel Jolley} {and}
  \bibinfo{person}{Karen~M Douglas}.} \bibinfo{year}{2014}\natexlab{}.
\newblock \showarticletitle{The Effects of Anti-Vaccine Conspiracy Theories on
  Vaccination Intentions}.
\newblock \bibinfo{journal}{\emph{Public Library of Science}}
  \bibinfo{volume}{9} (\bibinfo{year}{2014}).
\newblock
\urldef\tempurl%
\url{https://doi.org/10.1371/journal.pone.0089177}
\showDOI{\tempurl}


\bibitem[\protect\citeauthoryear{Kotzias, Bilge, and Caballero}{Kotzias
  et~al\mbox{.}}{2016}]%
        {kotzias16}
\bibfield{author}{\bibinfo{person}{Platon Kotzias}, \bibinfo{person}{Leyla
  Bilge}, {and} \bibinfo{person}{Juan Caballero}.}
  \bibinfo{year}{2016}\natexlab{}.
\newblock \showarticletitle{Measuring {PUP} Prevalence and {PUP} Distribution
  through Pay-Per-Install Services}. In \bibinfo{booktitle}{\emph{25th {USENIX}
  Security Symposium ({USENIX} Security 16)}}. \bibinfo{publisher}{{USENIX}
  Association}, \bibinfo{address}{Austin, TX}, \bibinfo{pages}{739--756}.
\newblock
\showISBNx{978-1-931971-32-4}
\urldef\tempurl%
\url{https://www.usenix.org/conference/usenixsecurity16/technical-sessions/presentation/kotzias}
\showURL{%
\tempurl}


\bibitem[\protect\citeauthoryear{Kouzy, Abi~Jaoude, Kraitem, El~Alam, Karam,
  Adib, Zarka, Traboulsi, Akl, and Baddour}{Kouzy et~al\mbox{.}}{2020}]%
        {kouzy20}
\bibfield{author}{\bibinfo{person}{Ramez Kouzy}, \bibinfo{person}{Joseph
  Abi~Jaoude}, \bibinfo{person}{Afif Kraitem}, \bibinfo{person}{Molly El~Alam},
  \bibinfo{person}{Basil Karam}, \bibinfo{person}{Elio Adib},
  \bibinfo{person}{Jabra Zarka}, \bibinfo{person}{Cindy Traboulsi},
  \bibinfo{person}{Elie Akl}, {and} \bibinfo{person}{Khalil Baddour}.}
  \bibinfo{year}{2020}\natexlab{}.
\newblock \showarticletitle{Coronavirus Goes Viral: Quantifying the COVID-19
  Misinformation Epidemic on Twitter}.
\newblock \bibinfo{journal}{\emph{Cureus}}  \bibinfo{volume}{12}
  (\bibinfo{date}{03} \bibinfo{year}{2020}).
\newblock
\urldef\tempurl%
\url{https://doi.org/10.7759/cureus.7255}
\showDOI{\tempurl}


\bibitem[\protect\citeauthoryear{Kumar and Shah}{Kumar and Shah}{2018}]%
        {Kumar18}
\bibfield{author}{\bibinfo{person}{Srijan Kumar} {and} \bibinfo{person}{Neil
  Shah}.} \bibinfo{year}{2018}\natexlab{}.
\newblock \showarticletitle{False Information on Web and Social Media: A
  Survey}.
\newblock \bibinfo{journal}{\emph{ArXiv}}  \bibinfo{volume}{abs/1804.08559}
  (\bibinfo{year}{2018}).
\newblock


\bibitem[\protect\citeauthoryear{Lantian, Bagneux, Delouvée, and
  Gauvrit}{Lantian et~al\mbox{.}}{2021}]%
        {criticalthinking}
\bibfield{author}{\bibinfo{person}{Anthony Lantian}, \bibinfo{person}{Virginie
  Bagneux}, \bibinfo{person}{Sylvain Delouvée}, {and} \bibinfo{person}{Nicolas
  Gauvrit}.} \bibinfo{year}{2021}\natexlab{}.
\newblock \showarticletitle{Maybe a free thinker but not a critical one: High
  conspiracy belief is associated with low critical thinking ability}.
\newblock \bibinfo{journal}{\emph{Applied Cognitive Psychology}}
  \bibinfo{volume}{35}, \bibinfo{number}{3} (\bibinfo{year}{2021}),
  \bibinfo{pages}{674--684}.
\newblock
\urldef\tempurl%
\url{https://doi.org/10.1002/acp.3790}
\showDOI{\tempurl}


\bibitem[\protect\citeauthoryear{Ledwich and Zaitsev}{Ledwich and
  Zaitsev}{2020}]%
        {ledwichzaitsev20}
\bibfield{author}{\bibinfo{person}{Mark Ledwich} {and} \bibinfo{person}{Anna
  Zaitsev}.} \bibinfo{year}{2020}\natexlab{}.
\newblock \showarticletitle{Algorithmic extremism: Examining YouTube's rabbit
  hole of radicalization}.
\newblock \bibinfo{journal}{\emph{First Monday}} (\bibinfo{date}{02}
  \bibinfo{year}{2020}).
\newblock
\urldef\tempurl%
\url{https://doi.org/10.5210/fm.v25i3.10419}
\showDOI{\tempurl}


\bibitem[\protect\citeauthoryear{Mai and Kulkarni}{Mai and Kulkarni}{2018}]%
        {mai18}
\bibfield{author}{\bibinfo{person}{Bin Mai} {and} \bibinfo{person}{Shailesh~S.
  Kulkarni}.} \bibinfo{year}{2018}\natexlab{}.
\newblock \showarticletitle{When Hackers Err: The Impacts of False Positives on
  Information Security Games}.
\newblock \bibinfo{journal}{\emph{Decis. Anal.}}  \bibinfo{volume}{15}
  (\bibinfo{year}{2018}), \bibinfo{pages}{90--109}.
\newblock


\bibitem[\protect\citeauthoryear{Mathur, Narayanan, and Chetty}{Mathur
  et~al\mbox{.}}{2018}]%
        {affiliateMarketingDisclosureMeasurement}
\bibfield{author}{\bibinfo{person}{Arunesh Mathur}, \bibinfo{person}{Arvind
  Narayanan}, {and} \bibinfo{person}{Marshini Chetty}.}
  \bibinfo{year}{2018}\natexlab{}.
\newblock \showarticletitle{Endorsements on Social Media: An Empirical Study of
  Affiliate Marketing Disclosures on YouTube and Pinterest}.
\newblock \bibinfo{journal}{\emph{Proc. ACM Hum.-Comput. Interact.}}
  \bibinfo{volume}{2}, \bibinfo{number}{CSCW}, Article \bibinfo{articleno}{119}
  (\bibinfo{date}{Nov.} \bibinfo{year}{2018}), \bibinfo{numpages}{26}~pages.
\newblock
\urldef\tempurl%
\url{https://doi.org/10.1145/3274388}
\showDOI{\tempurl}


\bibitem[\protect\citeauthoryear{Merril and Allen}{Merril and Allen}{2020}]%
        {insurancescamspp}
\bibfield{author}{\bibinfo{person}{Jeremy~B Merril} {and}
  \bibinfo{person}{Marshall Allen}.} \bibinfo{year}{2020}\natexlab{}.
\newblock \showarticletitle{“Trumpcare” Does Not Exist. Nevertheless
  Facebook and Google Cash In on Misleading Ads for “Garbage” Health
  Insurance.}
\newblock  (\bibinfo{year}{2020}).
\newblock
\urldef\tempurl%
\url{https://www.propublica.org/article/trumpcare-does-not-exist-nevertheless-facebook-and-google-cash-in-on-misleading-ads-for-garbage-health-insurance}
\showURL{%
\tempurl}


\bibitem[\protect\citeauthoryear{Mueller}{Mueller}{2019}]%
        {mueller19}
\bibfield{author}{\bibinfo{person}{Robert~S Mueller}.}
  \bibinfo{year}{2019}\natexlab{}.
\newblock \bibinfo{booktitle}{\emph{Report On The Investigation Into Russian
  Interference In The 2016 Presidential Election}}. Vol.~\bibinfo{volume}{1}.
\newblock


\bibitem[\protect\citeauthoryear{Nelson}{Nelson}{2020}]%
        {Nelsonadsense}
\bibfield{author}{\bibinfo{person}{Angie Nelson}.}
  \bibinfo{year}{2020}\natexlab{}.
\newblock \bibinfo{booktitle}{\emph{8 Ways to Monetize YouTube Videos (even
  without 4,000 watch hours)}}.
\newblock
\urldef\tempurl%
\url{https://theworkathomewife.com/monetize-youtube/}
\showURL{%
\tempurl}


\bibitem[\protect\citeauthoryear{Olshansky}{Olshansky}{2018}]%
        {olshansky18}
\bibfield{author}{\bibinfo{person}{Alex Olshansky}.}
  \bibinfo{year}{2018}\natexlab{}.
\newblock \emph{\bibinfo{title}{Conspiracy Theorizing and Religious Motivated
  Reasoning: Why the Earth ‘Must’ Be Flat}}.
\newblock \bibinfo{thesistype}{Master's\ thesis}. \bibinfo{school}{Texas Tech
  University}.
\newblock


\bibitem[\protect\citeauthoryear{Paolillo}{Paolillo}{2018}]%
        {Paolillo18}
\bibfield{author}{\bibinfo{person}{John~C. Paolillo}.}
  \bibinfo{year}{2018}\natexlab{}.
\newblock \showarticletitle{The Flat Earth phenomenon on YouTube}.
\newblock \bibinfo{journal}{\emph{First Monday}} \bibinfo{volume}{23},
  \bibinfo{number}{12} (\bibinfo{date}{Dec.} \bibinfo{year}{2018}).
\newblock
\urldef\tempurl%
\url{https://doi.org/10.5210/fm.v23i12.8251}
\showDOI{\tempurl}


\bibitem[\protect\citeauthoryear{Papadamou, Zannettou, Blackburn, Cristofaro,
  Stringhini, and Sirivianos}{Papadamou et~al\mbox{.}}{2021}]%
        {papadamou21}
\bibfield{author}{\bibinfo{person}{Kostantinos Papadamou},
  \bibinfo{person}{Savvas Zannettou}, \bibinfo{person}{Jeremy Blackburn},
  \bibinfo{person}{Emiliano~De Cristofaro}, \bibinfo{person}{Gianluca
  Stringhini}, {and} \bibinfo{person}{Michael Sirivianos}.}
  \bibinfo{year}{2021}\natexlab{}.
\newblock \showarticletitle{"How over is it?" Understanding the Incel Community
  on YouTube}.
\newblock \bibinfo{journal}{\emph{Proceedings of the ACM on Human-Computer
  Interaction}}  \bibinfo{volume}{5} (\bibinfo{year}{2021}), \bibinfo{pages}{1
  -- 25}.
\newblock


\bibitem[\protect\citeauthoryear{Papasavva, Blackburn, Stringhini, Zannettou,
  and Cristofaro}{Papasavva et~al\mbox{.}}{2021}]%
        {PapasavvaVoat}
\bibfield{author}{\bibinfo{person}{Antonis Papasavva}, \bibinfo{person}{Jeremy
  Blackburn}, \bibinfo{person}{Gianluca Stringhini}, \bibinfo{person}{Savvas
  Zannettou}, {and} \bibinfo{person}{Emiliano~De Cristofaro}.}
  \bibinfo{year}{2021}\natexlab{}.
\newblock \showarticletitle{“Is It a Qoincidence?”: An Exploratory Study of
  QAnon on Voat}. In \bibinfo{booktitle}{\emph{Proceedings of the Web
  Conference 2021}} (Ljubljana, Slovenia) \emph{(\bibinfo{series}{WWW '21})}.
  \bibinfo{publisher}{Association for Computing Machinery},
  \bibinfo{address}{New York, NY, USA}, \bibinfo{pages}{460–471}.
\newblock
\showISBNx{9781450383127}
\urldef\tempurl%
\url{https://doi.org/10.1145/3442381.3450036}
\showDOI{\tempurl}


\bibitem[\protect\citeauthoryear{Rasmussen}{Rasmussen}{2018}]%
        {rasmussen18}
\bibfield{author}{\bibinfo{person}{Leslie~Lynn Rasmussen}.}
  \bibinfo{year}{2018}\natexlab{}.
\newblock \showarticletitle{Parasocial Interaction in the Digital Age: An
  Examination of Relationship Building and the Effectiveness of YouTube
  Celebrities}.
\newblock \bibinfo{journal}{\emph{Social media and society}}
  \bibinfo{volume}{7} (\bibinfo{year}{2018}), \bibinfo{pages}{280--294}.
\newblock


\bibitem[\protect\citeauthoryear{Ribeiro, Ottoni, West, Almeida, and
  Meira}{Ribeiro et~al\mbox{.}}{2020}]%
        {ribeiropathways}
\bibfield{author}{\bibinfo{person}{Manoel~Horta Ribeiro},
  \bibinfo{person}{Raphael Ottoni}, \bibinfo{person}{Robert West},
  \bibinfo{person}{Virg\'{\i}lio A.~F. Almeida}, {and} \bibinfo{person}{Wagner
  Meira}.} \bibinfo{year}{2020}\natexlab{}.
\newblock \showarticletitle{Auditing Radicalization Pathways on YouTube}. In
  \bibinfo{booktitle}{\emph{Proceedings of the 2020 Conference on Fairness,
  Accountability, and Transparency}} (Barcelona, Spain)
  \emph{(\bibinfo{series}{FAT* '20})}. \bibinfo{publisher}{Association for
  Computing Machinery}, \bibinfo{address}{New York, NY, USA},
  \bibinfo{pages}{131–141}.
\newblock
\showISBNx{9781450369367}
\urldef\tempurl%
\url{https://doi.org/10.1145/3351095.3372879}
\showDOI{\tempurl}


\bibitem[\protect\citeauthoryear{Ridout, Franklin~Fowler, and
  Branstetter}{Ridout et~al\mbox{.}}{2010}]%
        {ridout10}
\bibfield{author}{\bibinfo{person}{Travis Ridout}, \bibinfo{person}{Erika
  Franklin~Fowler}, {and} \bibinfo{person}{John Branstetter}.}
  \bibinfo{year}{2010}\natexlab{}.
\newblock \showarticletitle{Political Advertising in the 21st Century: The Rise
  of the YouTube Ad}.
\newblock \bibinfo{journal}{\emph{American Political Science Association}}
  (\bibinfo{date}{08} \bibinfo{year}{2010}).
\newblock


\bibitem[\protect\citeauthoryear{Rodriguez}{Rodriguez}{2017}]%
        {rodriguez17}
\bibfield{author}{\bibinfo{person}{Paula~R Rodriguez}.}
  \bibinfo{year}{2017}\natexlab{}.
\newblock \emph{\bibinfo{title}{Effectiveness of YouTube Advertising: A Study
  of Audience Analysis Analysis}}.
\newblock \bibinfo{thesistype}{Master's\ thesis}. \bibinfo{school}{Rochester
  Institute of Technology}.
\newblock


\bibitem[\protect\citeauthoryear{Roozenbeek, Schneider, Dryhurst, Kerr,
  Freeman, Recchia, van~der Bles, and van~der Linden}{Roozenbeek
  et~al\mbox{.}}{2020}]%
        {roozenbeek20}
\bibfield{author}{\bibinfo{person}{Jon Roozenbeek}, \bibinfo{person}{Claudia
  Schneider}, \bibinfo{person}{Sarah Dryhurst}, \bibinfo{person}{John Kerr},
  \bibinfo{person}{Alexandra Freeman}, \bibinfo{person}{Gabriel Recchia},
  \bibinfo{person}{Anne~Marthe van~der Bles}, {and} \bibinfo{person}{Sander
  van~der Linden}.} \bibinfo{year}{2020}\natexlab{}.
\newblock \showarticletitle{Susceptibility to misinformation about COVID-19
  around the world}.
\newblock \bibinfo{journal}{\emph{Royal Society Open Science}}
  \bibinfo{volume}{7} (\bibinfo{date}{10} \bibinfo{year}{2020}).
\newblock
\urldef\tempurl%
\url{https://doi.org/10.1098/rsos.201199}
\showDOI{\tempurl}


\bibitem[\protect\citeauthoryear{Rubin, Bruggeman, and Steakin}{Rubin
  et~al\mbox{.}}{[n.\,d.]}]%
        {qanonsiege}
\bibfield{author}{\bibinfo{person}{Olivia Rubin}, \bibinfo{person}{Lucien
  Bruggeman}, {and} \bibinfo{person}{Will Steakin}.}
  \bibinfo{year}{[n.\,d.]}\natexlab{}.
\newblock \showarticletitle{QAnon emerges as recurring theme of criminal cases
  tied to US Capitol siege}.
\newblock \bibinfo{journal}{\emph{ABC News}} (\bibinfo{year}{[n.\,d.]}).
\newblock
\urldef\tempurl%
\url{https://abcnews.go.com/US/qanon-emerges-recurring-theme-criminal-cases-tied-us/story?id=75347445}
\showURL{%
\tempurl}


\bibitem[\protect\citeauthoryear{Sood and Enbody}{Sood and Enbody}{2011}]%
        {sood11}
\bibfield{author}{\bibinfo{person}{Aditya~K Sood} {and}
  \bibinfo{person}{Richard~J Enbody}.} \bibinfo{year}{2011}\natexlab{}.
\newblock \showarticletitle{Malvertising – exploiting web advertising}.
\newblock \bibinfo{journal}{\emph{Computer Fraud \& Security}}
  \bibinfo{volume}{2011}, \bibinfo{number}{4} (\bibinfo{year}{2011}),
  \bibinfo{pages}{11--16}.
\newblock
\showISSN{1361-3723}
\urldef\tempurl%
\url{https://doi.org/10.1016/S1361-3723(11)70041-0}
\showDOI{\tempurl}


\bibitem[\protect\citeauthoryear{Swart, Lopez, Mathur, and Chetty}{Swart
  et~al\mbox{.}}{2020}]%
        {affiliateMarketingBrowserExtension}
\bibfield{author}{\bibinfo{person}{Michael Swart}, \bibinfo{person}{Ylana
  Lopez}, \bibinfo{person}{Arunesh Mathur}, {and} \bibinfo{person}{Marshini
  Chetty}.} \bibinfo{year}{2020}\natexlab{}.
\newblock \showarticletitle{Is This An Ad?: Automatically Disclosing Online
  Endorsements On YouTube With AdIntuition}. In
  \bibinfo{booktitle}{\emph{Proceedings of the 2020 CHI Conference on Human
  Factors in Computing Systems}} (Honolulu, HI, USA)
  \emph{(\bibinfo{series}{CHI '20})}. \bibinfo{publisher}{Association for
  Computing Machinery}, \bibinfo{address}{New York, NY, USA},
  \bibinfo{pages}{1–12}.
\newblock
\showISBNx{9781450367080}
\urldef\tempurl%
\url{https://doi.org/10.1145/3313831.3376178}
\showDOI{\tempurl}


\bibitem[\protect\citeauthoryear{Tan, Ng, Omar, and Karupaiah}{Tan
  et~al\mbox{.}}{2018}]%
        {tan18}
\bibfield{author}{\bibinfo{person}{LeeAnn Tan}, \bibinfo{person}{See~Hoe Ng},
  \bibinfo{person}{Azahadi Omar}, {and} \bibinfo{person}{Tilakavati
  Karupaiah}.} \bibinfo{year}{2018}\natexlab{}.
\newblock \showarticletitle{What's on YouTube? A Case Study on Food and
  Beverage Advertising in Videos Targeted at Children on Social Media.}
\newblock \bibinfo{journal}{\emph{Childhood obesity}} \bibinfo{volume}{14},
  \bibinfo{number}{5} (\bibinfo{year}{2018}), \bibinfo{pages}{280--290}.
\newblock
\urldef\tempurl%
\url{https://doi.org/10.1089/chi.2018.0037}
\showDOI{\tempurl}


\bibitem[\protect\citeauthoryear{Thomas, Bursztein, Grier, Ho, Jagpal,
  Kapravelos, Mccoy, Nappa, Paxson, Pearce, Provos, and Rajab}{Thomas
  et~al\mbox{.}}{2015}]%
        {thomas15}
\bibfield{author}{\bibinfo{person}{Kurt Thomas}, \bibinfo{person}{Elie
  Bursztein}, \bibinfo{person}{Chris Grier}, \bibinfo{person}{Grant Ho},
  \bibinfo{person}{Nav Jagpal}, \bibinfo{person}{Alexandros Kapravelos},
  \bibinfo{person}{Damon Mccoy}, \bibinfo{person}{Antonio Nappa},
  \bibinfo{person}{Vern Paxson}, \bibinfo{person}{Paul Pearce},
  \bibinfo{person}{Niels Provos}, {and} \bibinfo{person}{Moheeb~Abu Rajab}.}
  \bibinfo{year}{2015}\natexlab{}.
\newblock \showarticletitle{Ad Injection at Scale: Assessing Deceptive
  Advertisement Modifications}. In \bibinfo{booktitle}{\emph{2015 IEEE
  Symposium on Security and Privacy}}. \bibinfo{pages}{151--167}.
\newblock
\urldef\tempurl%
\url{https://doi.org/10.1109/SP.2015.17}
\showDOI{\tempurl}


\bibitem[\protect\citeauthoryear{Thomas, Crespo, Rasti, Picod, Phillips,
  Decoste, Sharp, Tirelo, Tofigh, Courteau, Ballard, Shield, Jagpal, Rajab,
  Mavrommatis, Provos, Bursztein, and McCoy}{Thomas et~al\mbox{.}}{2016}]%
        {thomasppi}
\bibfield{author}{\bibinfo{person}{Kurt Thomas}, \bibinfo{person}{Juan
  A.~Elices Crespo}, \bibinfo{person}{Ryan Rasti}, \bibinfo{person}{Jean-Michel
  Picod}, \bibinfo{person}{Cait Phillips}, \bibinfo{person}{Marc-Andr{\'e}
  Decoste}, \bibinfo{person}{Chris Sharp}, \bibinfo{person}{Fabio Tirelo},
  \bibinfo{person}{Ali Tofigh}, \bibinfo{person}{Marc-Antoine Courteau},
  \bibinfo{person}{Lucas Ballard}, \bibinfo{person}{Robert Shield},
  \bibinfo{person}{Nav Jagpal}, \bibinfo{person}{Moheeb~Abu Rajab},
  \bibinfo{person}{Panayiotis Mavrommatis}, \bibinfo{person}{Niels Provos},
  \bibinfo{person}{Elie Bursztein}, {and} \bibinfo{person}{Damon McCoy}.}
  \bibinfo{year}{2016}\natexlab{}.
\newblock \showarticletitle{Investigating Commercial Pay-Per-Install and the
  Distribution of Unwanted Software}. In \bibinfo{booktitle}{\emph{25th
  {USENIX} Security Symposium ({USENIX} Security 16)}}.
  \bibinfo{publisher}{{USENIX} Association}, \bibinfo{address}{Austin, TX},
  \bibinfo{pages}{721--739}.
\newblock
\showISBNx{978-1-931971-32-4}
\urldef\tempurl%
\url{https://www.usenix.org/conference/usenixsecurity16/technical-sessions/presentation/thomas}
\showURL{%
\tempurl}


\bibitem[\protect\citeauthoryear{Tomlein, Pecher, Simko, Srba, M{\'o}ro,
  Stefancova, Kompan, Hrckova, Podrouzek, and Bielikov{\'a}}{Tomlein
  et~al\mbox{.}}{2021}]%
        {tomlein21}
\bibfield{author}{\bibinfo{person}{Mat{\'u}s Tomlein},
  \bibinfo{person}{Branislav Pecher}, \bibinfo{person}{Jakub Simko},
  \bibinfo{person}{Ivan Srba}, \bibinfo{person}{R{\'o}bert M{\'o}ro},
  \bibinfo{person}{Elena Stefancova}, \bibinfo{person}{Michal Kompan},
  \bibinfo{person}{Andrea Hrckova}, \bibinfo{person}{Juraj Podrouzek}, {and}
  \bibinfo{person}{M{\'a}ria Bielikov{\'a}}.} \bibinfo{year}{2021}\natexlab{}.
\newblock \showarticletitle{An Audit of Misinformation Filter Bubbles on
  YouTube: Bubble Bursting and Recent Behavior Changes}.
\newblock \bibinfo{journal}{\emph{Fifteenth ACM Conference on Recommender
  Systems}} (\bibinfo{year}{2021}).
\newblock


\bibitem[\protect\citeauthoryear{van Prooijen}{van Prooijen}{2019}]%
        {gullible}
\bibfield{author}{\bibinfo{person}{Jan~Willem van Prooijen}.}
  \bibinfo{year}{2019}\natexlab{}.
\newblock \bibinfo{booktitle}{\emph{Belief in Conspiracy Theories}}.
\newblock \bibinfo{pages}{319--332}.
\newblock
\showISBNx{9780429203787}
\urldef\tempurl%
\url{https://doi.org/10.4324/9780429203787-17}
\showDOI{\tempurl}


\bibitem[\protect\citeauthoryear{Vanwesenbeeck, Hudders, and
  Ponnet}{Vanwesenbeeck et~al\mbox{.}}{2020}]%
        {vanwesenbeeck20}
\bibfield{author}{\bibinfo{person}{Ini Vanwesenbeeck}, \bibinfo{person}{Liselot
  Hudders}, {and} \bibinfo{person}{Koen Ponnet}.}
  \bibinfo{year}{2020}\natexlab{}.
\newblock \showarticletitle{Understanding the YouTube Generation: How
  Preschoolers Process Television and YouTube Advertising}.
\newblock \bibinfo{journal}{\emph{Cyberpsychology, Behavior, and Social
  Networking}}  \bibinfo{volume}{23} (\bibinfo{date}{04} \bibinfo{year}{2020}).
\newblock
\urldef\tempurl%
\url{https://doi.org/10.1089/cyber.2019.0488}
\showDOI{\tempurl}


\bibitem[\protect\citeauthoryear{Vesnic-Alujevic and Bauwel}{Vesnic-Alujevic
  and Bauwel}{2014}]%
        {alujevic14}
\bibfield{author}{\bibinfo{person}{Lucia Vesnic-Alujevic} {and}
  \bibinfo{person}{Sofie~Van Bauwel}.} \bibinfo{year}{2014}\natexlab{}.
\newblock \showarticletitle{YouTube: A Political Advertising Tool? A Case Study
  of the Use of YouTube in the Campaign for the European Parliament Elections}.
\newblock \bibinfo{journal}{\emph{Journal of Political Marketing}}
  \bibinfo{volume}{13}, \bibinfo{number}{3} (\bibinfo{year}{2014}),
  \bibinfo{pages}{195--212}.
\newblock
\urldef\tempurl%
\url{https://doi.org/10.1080/15377857.2014.929886}
\showDOI{\tempurl}
\showeprint{https://doi.org/10.1080/15377857.2014.929886}


\bibitem[\protect\citeauthoryear{Xing, Meng, Lee, Weinsberg, Sheth, Perdisci,
  and Lee}{Xing et~al\mbox{.}}{2015}]%
        {xing15}
\bibfield{author}{\bibinfo{person}{Xinyu Xing}, \bibinfo{person}{Wei Meng},
  \bibinfo{person}{Byoungyoung Lee}, \bibinfo{person}{Udi Weinsberg},
  \bibinfo{person}{Anmol Sheth}, \bibinfo{person}{Roberto Perdisci}, {and}
  \bibinfo{person}{Wenke Lee}.} \bibinfo{year}{2015}\natexlab{}.
\newblock \showarticletitle{Understanding Malvertising Through Ad-Injecting
  Browser Extensions}. In \bibinfo{booktitle}{\emph{Proceedings of the 24th
  International Conference on World Wide Web}} (Florence, Italy)
  \emph{(\bibinfo{series}{WWW '15})}. \bibinfo{publisher}{International World
  Wide Web Conferences Steering Committee}, \bibinfo{address}{Republic and
  Canton of Geneva, CHE}, \bibinfo{pages}{1286–1295}.
\newblock
\showISBNx{9781450334693}
\urldef\tempurl%
\url{https://doi.org/10.1145/2736277.2741630}
\showDOI{\tempurl}


\bibitem[\protect\citeauthoryear{Zeng, Kohno, and Roesner}{Zeng
  et~al\mbox{.}}{2021}]%
        {badads}
\bibfield{author}{\bibinfo{person}{Eric Zeng}, \bibinfo{person}{Tadayoshi
  Kohno}, {and} \bibinfo{person}{Franziska Roesner}.}
  \bibinfo{year}{2021}\natexlab{}.
\newblock \bibinfo{booktitle}{\emph{What Makes a “Bad” Ad? User Perceptions
  of Problematic Online Advertising}}.
\newblock \bibinfo{publisher}{Association for Computing Machinery},
  \bibinfo{address}{New York, NY, USA}.
\newblock
\showISBNx{9781450380966}
\urldef\tempurl%
\url{https://doi.org/10.1145/3411764.3445459}
\showURL{%
\tempurl}


\bibitem[\protect\citeauthoryear{Zummeren and Ballard}{Zummeren and
  Ballard}{2021}]%
        {raditube}
\bibfield{author}{\bibinfo{person}{Erik~Van Zummeren} {and}
  \bibinfo{person}{Cameron Ballard}.} \bibinfo{year}{2021}\natexlab{}.
\newblock \bibinfo{booktitle}{\emph{About Raditube}}.
\newblock
\urldef\tempurl%
\url{https://extension.raditube.com/about}
\showURL{%
\tempurl}


\end{thebibliography}

\balance
\newpage
\appendix
\section{Appendix}
\label{sec:appendix}

A full list of videos and channels we crawled, along with demonetization status during our period of observation can be found at: https://github.com/Camq543/YoutubeAds 

\begin{figure}[h]
    \centering
    \includegraphics[width=\columnwidth]{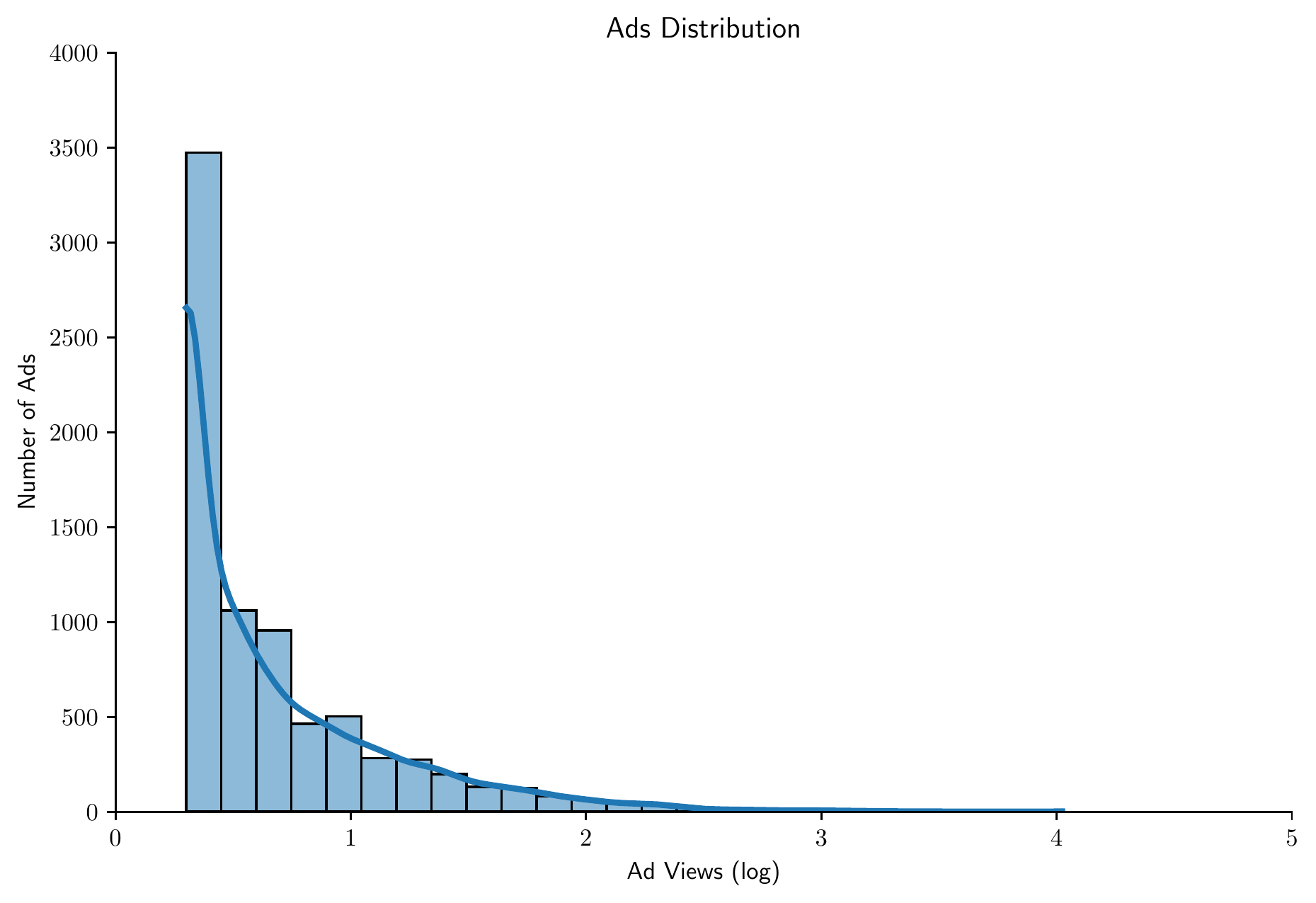}
    \caption{Histogram of the number of impressions for unique advertisers. More than 40\% of advertisers were only seen once, while two were seen more than 20,000 times. X-axis uses log base 10.}
    \label{fig:adshistogram}
\end{figure}

\begin{figure}[h]
    \centering
    \includegraphics[width=\columnwidth]{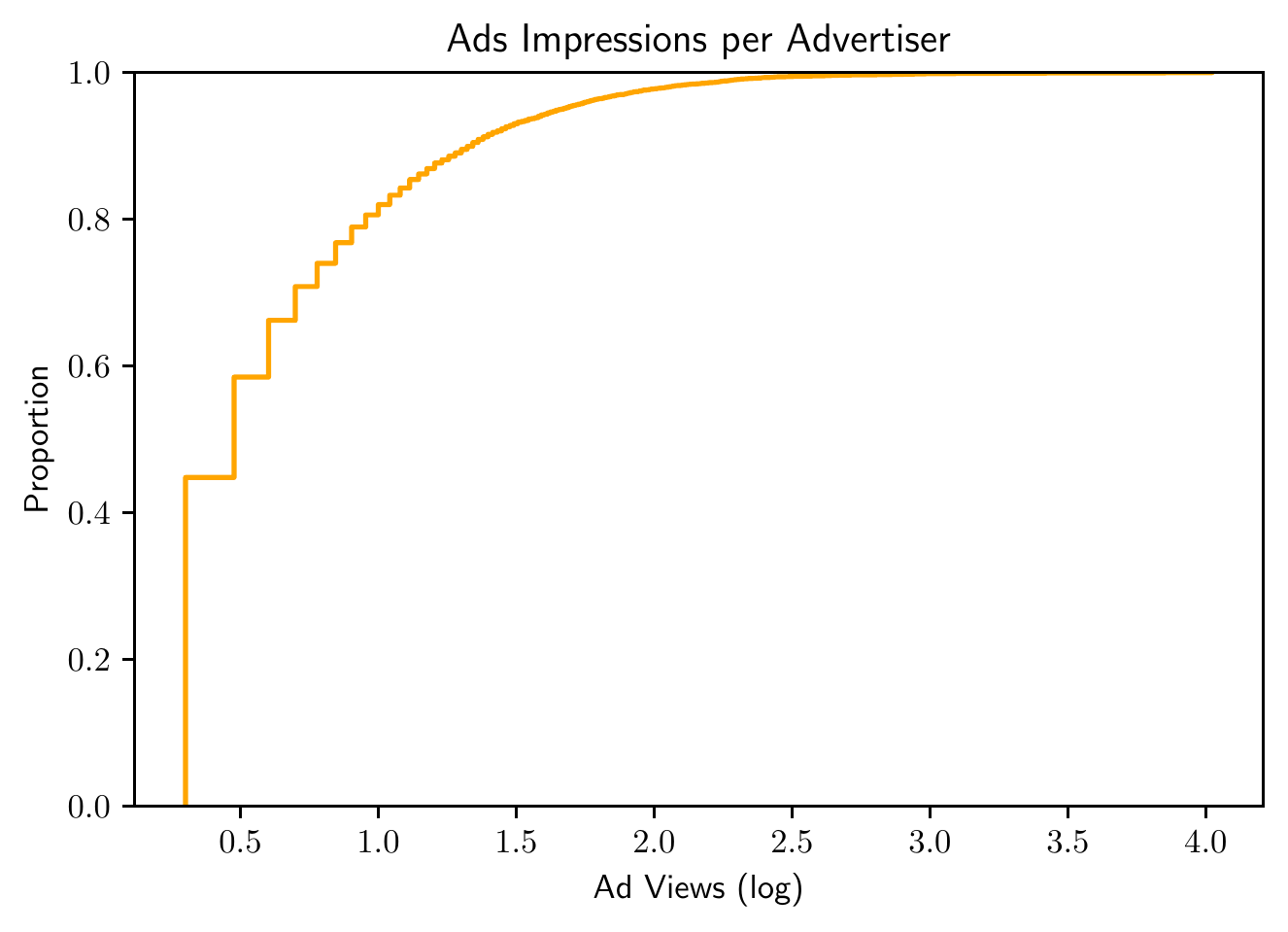}
    \caption{CDF of the number of impressions for unique advertisers observed. X-axis uses log base 10.}
    \label{fig:adscdf}
\end{figure}

\begin{table*}[p]
\begin{tabular}{lp{8cm}l}
\toprule
Category &   Category Explanation &  Example\\
Business Type           &   & \\
\midrule
Aggregator Sites        &       Link aggregators, review sites, and other aggregated information advertising for products not owned by them &  poki[.com] \\
\addlinespace
Information Media       &       News and review websites not advertising specific products &         wired[.com] \\
\addlinespace
Merchant                &       Products or services that do not fall under self improvement &         products.4patriots[.com] \\
\addlinespace
Self Improvement        &       Education and self-help; they are selling a better life &         theshiftnetwork[.com] \\
\addlinespace
Special Interest Groups &       Charities, Governmental or Political organizations, non-profits  &         www.boldpac[.com] \\
    &   & \\
\midrule
Content Type                   &                 &               \\
\midrule
Alternative Health       &      Claims to address a specific health problem or problems and is not widely accepted by in western medicine. Natural supplements, crystal/energy healing, essential oils, etc.  &         www.bookofremedies[.org] \\
\addlinespace
Beauty                   &      Cosmetics, grooming, etc. &         www.manscaped[.com] \\
\addlinespace
Business                 &      Advertised to help with business or marketing, and/or improve one's success as a business person & growthcave[.com] \\
\addlinespace
Education                &      Formalized educational program in any area &         bardacademy.simons-rock[.edu] \\
\addlinespace
Electronics              &      Having to do with physical electronics &         futureelectronics[.com] \\
\addlinespace
Entertainment            &      Music, movies, toys, etc. Does not include games &         www.thesuicidesquad[.com] \\
\addlinespace
Fashion                  &      Clothes, jewelry, accessories, etc. &         www.prada[.com] \\
\addlinespace
Financial                &      Having to do with personal investment and financial health &         www.interactivebrokers[.com] \\
\addlinespace
Food and Drink           &      Edible goods and food delivery. Does not include products marketed as health products. &         www.smuckers[.com] \\
\addlinespace
Games                    &      Games and gaming related content. &         hero-wars[.com] \\
\addlinespace
Gold and Precious Metals &      Having to do with precious metals collecting or investing &         www.moneymetals[.com] \\
\addlinespace
Government               &      Official government sources  &         www.marines[.com] \\
\addlinespace
Home Goods               &      Furniture, cooking and cleaning supplies, hobbies, etc. &         www.dyson[.com] \\
\addlinespace
Industrial               &      Advertised for use in commercial industry, not hobbyist tools &         channellock[.com] \\
\addlinespace
Insurance                &      For or about insurance or insurance-like services &         www.statefarm[.com] \\
\addlinespace
Lifestyle                &      Claims to change lifestlye in ways unrelated to specific needs or medical conditions. Fitness, dating, actualization, etc.  &         www.noom[.com] \\
\addlinespace
Major Retailer           &      Any large-scale retailer with many types of products &         walmart[.com] \\
\addlinespace
Medical                  &      Marketed to address health problems or to be used in healthcare and is widely accepted in western medicine. Prescription drugs, FDA approved supplements, medical machines, etc. &         honeybeepharmacy[.com] \\
\addlinespace
Political                &      Explicitly involving politics and not an official governmental source. Campaigns, PACs, issue ads, advocacy groups, etc.  &         www.judicialwatch[.org] \\
\addlinespace
Software                 &      Any product or service that is software or relies on it to work and is not encompassed by another category. This includes individual software as well as SaaS, IaaS, PaaS, etc.  &         argoskyc[.com] \\
\addlinespace
Transportation           &      Land, air, and sea transportation, vehicles, and accessories &         www.toyotires[.com] \\
\bottomrule
\end{tabular}
\caption{Detailed explanation of labeling categories, and examples found in the data.}
\label{tab:catsubcat}
\end{table*}

\end{document}